\documentclass[useAMS,usenatbib]{mn2e}
\usepackage{amssymb,amsmath,graphicx}
\usepackage{textcomp}
\usepackage{graphics}
\usepackage{longtable}[1]
\usepackage{lscape}
\usepackage{color}
\usepackage{xcolor}
\usepackage{rotating}
\usepackage{cite}
\usepackage{ulem}

\usepackage{lastpage}
\usepackage{journal_names}
\usepackage{multicol}
\usepackage{multirow}
\title[SSC LFs in LIRGs]{The K-band luminosity functions of super star clusters in luminous infrared galaxies, their slopes, and the effects of blending}

\author[Z. Randriamanakoto et al.]{Z. Randriamanakoto$^{1,2}$\thanks{E-mail:
zara@saao.ac.za}, P. V\"ais\"anen$^{1,3}$, S. Ryder$^{4}$, E. Kankare$^{5}$, J. Kotilainen$^{6}$,  \newauthor  S. Mattila$^{5,6}$ \\
$^{1}$South African Astronomical Observatory, P.O. Box 9 Observatory, Cape Town, South Africa\\
$^{2}$University of Cape Town, Astronomy Department, Private Bag X3, Rondebosch 7701, South Africa\\
$^{3}$Southern African Large Telescope, P.O. Box 9 Observatory, Cape Town, South Africa\\
$^{4}$Australian Astronomical Observatory, P.O. Box 915, North Ryde, NSW 1670, Australia\\
$^{5}$Tuorla Observatory, Department of Physics and Astronomy, University of Turku, V\"ais\"al\"antie 20, FI-21500 Piikki\"{o}, Finland\\
$^{6}$Finnish Centre for Astronomy with ESO (FINCA), University of Turku, V\"ais\"al\"antie 20, FI-21500 Piikki\"o, Finland
}

\begin{document}
\normalem%{\em phasized\/} and \emph{asized} In LaTeX, by default, these are underlined; use \normalem or [normalem] to restore italics
\date{Accepted 2013 January 29. Received 2013 January 29; in original form 2012 September 6}

\pagerange{\pageref{firstpage}--\pageref{LastPage}} 
\pubyear{2012}

\maketitle

\label{firstpage}

\begin{abstract}

Super star clusters (SSCs) are typically found in interacting galaxies and trace an extreme form of star-formation.  We present a $K$-band study of SSC candidates in a sample of local luminous infrared galaxies (LIRGs) using two adaptive optics instruments (VLT/NACO and Gemini/ALTAIR/NIRI).  In addition to facilitating SSC detections in obscured environments, this work introduces SSC studies in hosts with higher star-formation rates (SFRs) than most previous studies.  We find that the luminosity functions (LFs) of the clusters are reasonably well-fitted by a single power-law with the values of the index $\alpha$ ranging between 1.5 to 2.4 with an average value of $\alpha \approx 1.9$. This value appears to be less steep than the average $\alpha \approx 2.2$ in normal spiral galaxies. Due to the host galaxy distances involved (median $D_L \sim 70$ Mpc) blending effects have to be taken into account, and are investigated using Monte Carlo simulations of blending effects for LFs and a photometric SSC analysis of the well-studied Antennae system which is artificially redshifted to distances of our sample.   While blending tends to flatten LFs our analyses show that $\Delta \alpha$ is less than $\sim$\,0.1 in our sample. The simulations also show that in the luminosity range, $M_K < -13$, considered in this work the extracted SSC luminosities are generally dominated by a single dominant star cluster rather than several knots of SF.  We present resolution- and distance-dependent SSC surface density confusion limits and show how  blending rates and aperture sizes affect the LFs.  The smallest possible apertures should be used in crowded regions.

\end{abstract}

\begin{keywords}
galaxies: interactions - galaxies: star clusters: general - infrared: galaxies  
\end{keywords}
\section{Introduction}

Young massive star clusters are related to triggers of star formation (SF) in galaxies and contain clues to the physical conditions under which extremely strong SF happens. These clusters, often called \textgravedbl super star clusters\textacutedbl (SSCs) when masses are in the $10^5$\,-\,$10^7$\,$M_{\odot}$ range, are typically found in interacting and merging gas-rich galaxies.  Their birth, evolution and disruption are not well understood while these issues are very interesting in the context of clustered SF in general \citep[e.g.][]{2003ARA&A..41...57L} and in particular since they might be the progenitors of globular clusters \citep[e.g.][]{1992ApJ...384...50A,1992AJ....103..691H, 1996ApJ...466L..83H, 1997ApJ...480..235E}.  For a recent review of young and massive star clusters see \citet{2010ARA&A..48..431P}.

SSCs can be used to trace the history of bursts of SF in galaxies \citep[e.g][]{2008MNRAS.390..759B,  2008ApJ...685L..31E, 2011ApJ...735...56E, 2011MNRAS.415.2388A} since  SSCs are bright and relatively simple to model as single stellar populations.   It is important to  understand how wide-spread starbursts can be in interactions and mergers, particularly in the context of high-redshift galaxy formation, since these conditions could be similar to what is happening at {\it z} $>$\,1. Local LIRGs (luminous IR galaxies,  $\log L_{IR} / L_{\odot} >  11$) exhibit extreme star formation and large populations of SSCs and they are believed to be good analogs for higher-$z$ star formation in general \citep[e.g.][]{2009ApJ...697..660A, 2011A&A...533A.119E, 2012MNRAS.419.1176T}.
  
Although the young and compact SSCs (age\,$\sim$\,10\,-\,100\,Myr, 
$\rm r_{eff}\sim$\,3\,-\,5\,pc) are often confined to starbursts and interacting systems 
\citep[e.g.][]{1999AJ....118.1551W} 
they can also form in more quiescent environments such as in the circum-nuclear star-forming rings (e.g. \citealp{1993AJ....105.1369B}), in nearby dwarf galaxies (\citealp{1998A&A...335...85O}) and even in normal spiral galaxies (\citealp{2002AJ....124.1393L}). The difference is just that the population of SSCs is much larger in galaxies exhibiting an extreme starburst environment with a violent SF.  Over the past decades, research has focused more on SSCs within nearby ($D_{L}$\,$\lesssim$\,25\,\rm \,Mpc) non-LIRG starburst systems, simply because it is much easier (e.g.  \citealp{1993AJ....106.1354W,1999AJ....118.1551W,2003A&A...397..473B,2008A&A...487..937H}). SSCs were, in fact, one of the first discoveries of the {\em Hubble Space Telescope (HST)}  while imaging the central regions of NGC 1275 \citep{1992AJ....103..691H}.  In particular the closest major merger system, the Antennae (NGC\,4038/4039, log\,$L_{IR}$\,=\,${11.0}$) has been amongst the most studied SSC hosts, providing a sample of thousands of these clusters.   There have been some individual cases studied further away, and at higher SF levels, such as 
 Haro\,11 with log\,$L_{IR}$\,=\,${11.22}$ at 81.2\,Mpc and the Bird galaxy (IRAS\,19115-2124) with log\,$L_{IR}$\,=\,${11.87}$ at 206\,Mpc, by  \citet{2010MNRAS.407..870A} and \citet{2008MNRAS.384..886V}, respectively. 
 The only published work thus far characterising SSCs from a significant sample of  galaxies is that of \citet{2011AJ....142...79M}, who analysed optically-selected star forming knots, SSCs, or complexes of clusters in 32 LIRGs and ultraluminous IR galaxies (ULIRGs, $\log L_{IR} / L_{\odot} > 12$).  Their results indicate for example that SSCs in post-mergers are larger, more luminous and redder, than in less advanced systems.  An even larger sample of SSCs in 87 LIRGs has been recently studied by \citet{2011PhDT.........8V}. 

 It is extremely important to study SSC characteristics over a wide range of galaxy properties and SF environments, given that there is a very lively debate on the universality of SSC mass functions  and disruption mechanisms \citep[see e.g.][]{2009Ap&SS.324..183L}. For example, \citet{2009ApJ...704..453F},  \citet{2007AJ....133.1067W, 2010AJ....140...75W} and \citet{2010ApJ...719..966C, 2011ApJ...727...88C} argue that any  turnover in the mass-function results from a constant disruption triggered by internal mechanisms (e.g. stellar evolution, 2-body relaxation) over the cluster lifetime.  In contrast, \citet{2005A&A...429..173L}, \citet{2006MNRAS.371..793G}, \citet{2008ASPC..388..279L}, \citet{2011MNRAS.417L...6B} and \citet{2012MNRAS.421.1927K} propose that the mass of the cluster and/or the environment of its host galaxy play an important role in the disruption model (low-mass clusters are less likely to survive the disruption, while environments with low giant molecular cloud densities will endure less disruption) that will later define  the cluster mass function (CMF) shape. 

The luminosity function (LF) of SSCs, which is suggested to be a reflection of the cluster initial mass function, is a common tool used to help understand their formation and evolution \citep[e.g.][]{1999AJ....118.1551W, 2006A&A...450..129G}.  While theoretical and observational studies have agreed that the LF of globular clusters is well-fitted with a Gaussian function, there are varying results regarding the LF of SSCs: though power-laws are usually found, and expected theoretically, log-normal distributions and bent power-laws  have also been advocated  (e.g. \citealp{1997ApJ...480..235E}, \citealp{1999A&A...342L..25F}, \citealp{2002AJ....124.1393L}, \citealp{2003ApJ...583L..17D}, \citealp{2003ARA&A..41...57L}, \citealp{2006A&A...450..129G}, \citealp{2007MNRAS.377...91A}, \citealp{2010AJ....140...75W}, \citealp{2011AJ....142...79M}).  To complicate matters, there are several observational issues to overcome as well to pin down the shapes, relating, for example, to stellar contamination, completeness, and blending (e.g. \citealp{2003A&A...409..523R, 2007MNRAS.377...91A, 2009ApJ...704..453F, 2010MNRAS.407..870A}). The  typical single power-law approximation (e.g. \citealp{1997ApJ...480..235E}) takes the form:
\begin{equation} 
N(L)dL \sim L^{-\alpha}dL,
\label{equone}
\end{equation}
where $\alpha$\,$\sim$\,2 with scatter in the range 1.5 to 2.5. \citet{2006A&A...446L...9G}, \citet{2008MNRAS.390..759B}, \citet{2008A&A...487..937H} and \citet{2009A&A...494..539L} among others, on the other hand, suggest a double power-law  or a Schechter function to better fit the cluster luminosity function (CLF) due to a turnover located at higher luminosities, which could be a sign of a truncated CMF. The interesting question is whether there are any systematic trends to those LF index values. 
 Normal spiral galaxies tend to have LFs biased to the steeper range of slopes at $\alpha \sim 2-2.4$ \citep[e.g.][]{2002AJ....124.1393L, 2006A&A...450..129G, 2009A&A...501..949M}.  Recent results from larger samples of LIRGs \citep{2011AJ....142...79M,2011PhDT.........8V} show evidence that $\alpha \approx 1.8-2.0$, though at distances $>200$\,Mpc where resolution and blending effects  may play a role, they find that the LF flattens to $\alpha \sim$\,1.7.  
Moreover,  \citet{2003A&A...408..887O} and \citet{2010MNRAS.407..870A, 2011MNRAS.414.1793A, 2011MNRAS.415.2388A} derived flatter slopes ranging between 1.5 to 1.8 for the $I$-band CLF in their very strongly star forming blue compact dwarf galaxies (54 to 82\,Mpc),  while \citet{1998ApJ...492..116S} and \citet{2010AJ....140...63I} find similar slopes between 1.1 and 1.8 in the $B$-band for three (U)LIRGs.

In our ongoing study we are expanding the SFR range of SSC host galaxies to include LIRGs from their lowest limit to ULIRGs, and including galaxies in various stages of interactions, from isolated and paired galaxies, to interacting, merging and merger remnant stages with a sample of (ultimately) dozens of targets \citep{2012arXiv1202.6236V}.   Furthermore, most of the studies thus far have been done in the optical, making extinction effects potentially difficult in the notoriously complex dusty environments of gas-rich interactions. We use $K_{S}$-band\footnote{Hereafter, we will refer to both the Johnson $K$-band and 2MASS $K_{S}$-band observations as $K$-band.} adaptive optics (AO) observations which match the spatial resolution of the $HST$ optical studies. Crucially, the use of near infrared (NIR) has great potential in opening a new angle into the SSC populations in that they probe deeper into the dusty birth regions of the SSCs and the very obscured regions in the inner parts of the galaxies. There is also an interesting time-window at ages  of $\sim$\,10\,Myr when SSCs are expected to be very NIR-luminous due to their high-mass stars entering the red supergiant (RSG) phase. There are very few NIR studies of SSCs compared with the optical ones, including \citet{1999A&A...351..834L}, \citet{2005A&A...443...41M}, \citet{2006ApJ...650..835A}, \citet{2007ApJ...660..288P}, \citet{2008MNRAS.384..886V}, and \citet{2010MNRAS.407..870A}. 

The main specific objectives of this work are to derive the $K$-band LFs of the massive star clusters and to evaluate the effect of blending on the LFs. In forthcoming papers we will study the mass functions and age distributions of the targets by combining optical data with the NIR photometry.  
 The paper is organised as follows: we describe the data and its reduction in $\S$\,\ref{reduction}. The cluster analysis is presented in $\S$\,3. The NIR CLFs of the sample are presented in $\S$\,\ref{fit_LF} and are interpreted in $\S$\,5; finally, we summarise our findings then suggest our future work in $\S$\,\ref{summary}.

Throughout the paper we assume the standard cosmology:
$H_0 = 73$ km s$^{-1}$ Mpc$^{-1}$, $\Omega_M = 0.27$, and
$\Omega_{\Lambda} = 0.73$.

\begin{table*}
  \centering
  \begin{tabular}{|l|c|c|c|c|c|c|c|c|c|c|}
  \hline
  Galaxy name& Exp time &RA &DEC &{\it l}&{\it b}&log~$L_{IR}$&SFR&$m - M$ &$D_{L}$  \\
             & ({\it sec}) & (J2000) & (J2000) & ($degrees$) & ($degrees$) & ($L_{\odot}$)&($M_{\odot}yr^{-1}$)&& (Mpc) \\
   \hline
   \hline

IC 694  & 1260 & 11 28 33.5 & +58 33 45& 141.9& 55.4& 11.60$^a$&67.7&33.28&45.3 \\
NGC 3690& 2192 &11 28 29.8 & +58 33 43 & 141.9& 55.4& 11.52$^a$&56.3&33.28& 45.3 \\
CGCG 049-057& 1680 & 15 13 13.1 & +07 13 32 & 8.9& 50.9&  11.27&31.6&33.76& 56.4 \\
IRAS F17578-0400& 1470 & 18 00 31.9 & $-$04 00 53 & 23.4& 9.4 &11.35&38.1&33.79& 57.3\\
IRAS F17138-1017& 990 & 17 16 35.8 & $-$10 20 39 & 12.2& 15.6 &11.42&44.7&34.29& 72.2 \\
IRAS 18293-3413& 1230 & 18 32 41.1 & $-$34 11 27 & 0.1& -11.3 & 11.81 &109.8&34.37& 74.6 \\
MGC +08-11-002& 1140 & 05 40 43.7 & +49 41 41& 161.6& 9.9  & 11.41&43.7&34.51& 79.9\\
IRAS F16516-0948& 900 & 16 54 24.0&$-$09 53 21 & 9.5& 20.5 & 11.24&29.5&34.88& 94.8\\
IC 883& 1440 & 13 20 35.3 & +34 08 22 & 82.9& 80.6& 11.67 &79.5&35.02& 101 \\
IRAS 19115-2124& 1410 &19 14 30.9 & $-$21 19 07 & 16.1 & -14.4& 11.87 &126&36.56& 206 \\

   \hline
\end{tabular}
  \caption{\small Total exposure times, equatorial (RA, DEC) and galactic ({\it l,\,b}) coordinates of the observed sample are shown in this table ordered by distance. The luminosity distance,  $D_{L}$,  and the distance modulus, $m - M$, are retrieved from NASA/IPAC Extragalactic Database (NED) while  the logarithmic value of the IR luminosity, $L_{IR}$, is as estimated by \citet{2003AJ....126.1607S} with a slightly different setting of the cosmological parameters.  \hspace{7mm} $^a$ These two galaxies are components of an interacting system Arp\,299; the total $L_{IR}$ is divided between IC\,694 (known as nucleus A) and NGC\,3690 (nuclei B+C) following \citet{2012ApJ...756..111M} the circum-nuclear SF portion being evenly split.  An empirical relation using $L_{IR}$ by \citet{1998ARA&A..36..189K} is used to derive the SFR.
 }
\label{coords_table}
\end{table*}

\begin{figure*}
\resizebox{0.9\hsize}{!}{\rotatebox{0}{\includegraphics{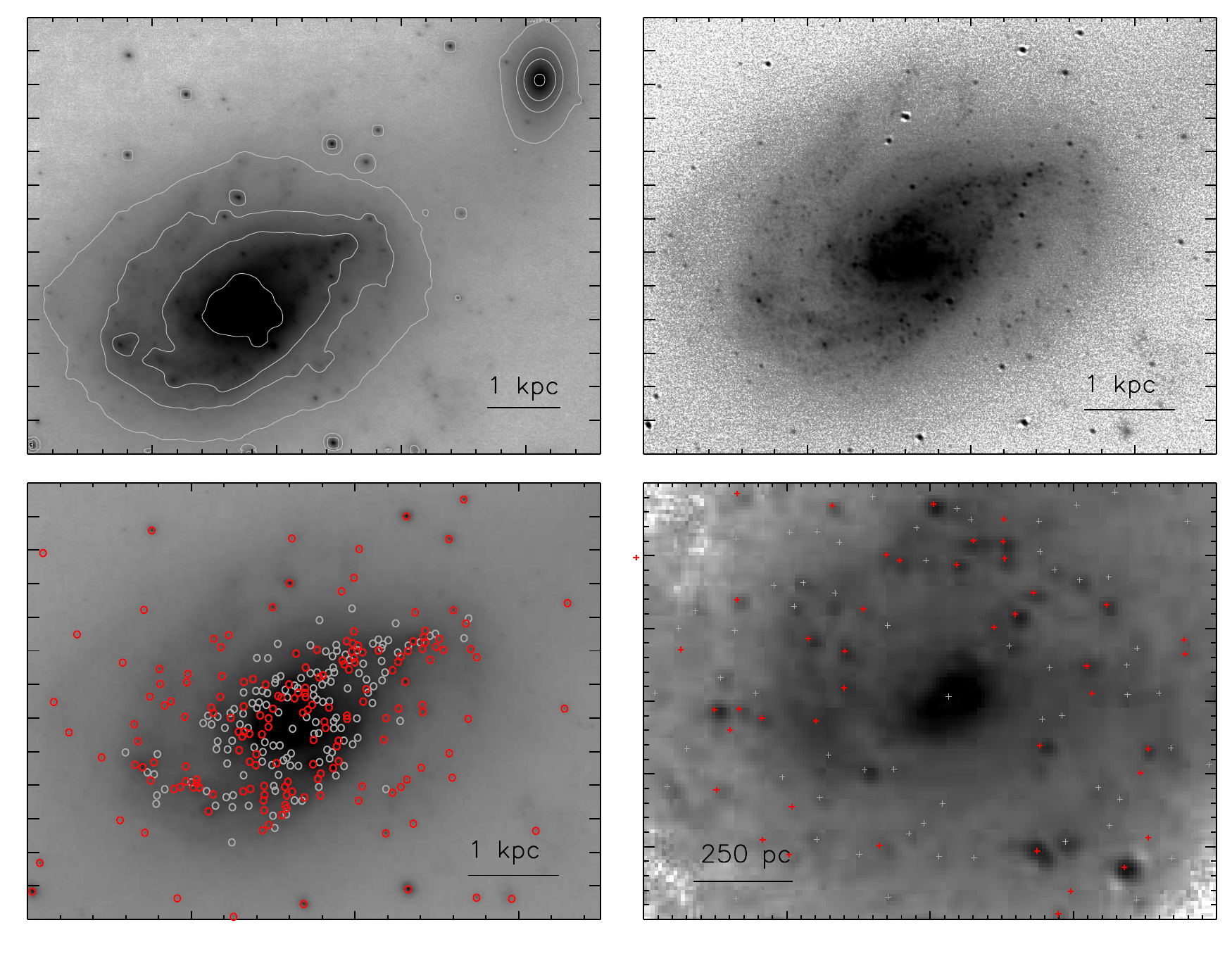}}}\\
\caption{\small The NACO field of IRAS\,18293-3413.  {\em Top left}:  The field of the whole interacting system with contours demarcating the four selected background regions for completeness analysis.  {\em Top-right}:  A slightly smaller field around the primary galaxy is shown after unsharp masking.  The SSC detections are made from this image. {\em Lower left}:  All SSC candidate detections are overlaid as white on the original image (where the photometry is performed) while those that meet all the SSC candidate selection criteria presented in Section~\ref{selection} are shown as red. {\em Lower-right}:  Same as previous, but only for a zoomed-in region around the nucleus. Small tick marks are in 1\textacutedbl\,units, except in the last panel where they are in 0.1\textacutedbl\,units.}
\label{SSC_18}
\end{figure*}

\section{Observations and data reduction}\label{reduction}

We analyse a representative sample of ten local LIRGs from an ongoing NIR AO survey which is mainly intended to search for dust-obscured core-collapse supernovae \citep[see e.g.][]{2007ApJ...659L...9M, 2008ApJ...689L..97K, 2012ApJ...744L..19K, 2009arXiv0908.3495V}. The targets analysed here were drawn from the {\it IRAS Revised Bright Galaxy Sample} \citep{2003AJ....126.1607S}. They lie at redshifts of 0.01\,$<$\,{\it z}\,$<$\,0.05, their luminosities range from $\log (L_{IR} / L_{\odot})=11.2$ to 11.9, and they were selected to be starburst-dominated based on their IRAS colours ($f_{25}/f_{60}$\,$<$\,0.2).   The observations were taken in $K$-band using two different ground-based instruments with AO imaging: the NAOS-CONICA on the ESO Very Large Telescope (VLT/NACO) and the ALTAIR/NIRI on the Gemini-North telescope (ALTAIR/NIRI).
For AO correction, natural guide stars were used for the NACO data, whereas laser guide stars, with a tip/tilt reference star, were used with the ALTAIR/NIRI data -- this requirement of suitably bright reference stars is a further sample selection constraint, but does not bias the LIRG characteristics in any way.

\subsection{NACO data}

Our present sample includes two NACO galaxies, IRAS\,18293-3413 and IRAS\,19115-2124, the latter also being the subject of our pilot SSC study in relatively distant galaxies compared to most SSC works \citep{2008MNRAS.384..886V}. The data taken with the VLT UT4 and NACO S27 camera have a plate scale of 0.027\textacutedbl$\rm pix^{-1}$ and a field of view (FOV) of 27\textacutedbl. The AO correction worked well resulting in a FWHM\,$\sim$\,0.1\textacutedbl~for the point sources. Frames were taken with exposure times of 30\,{\it sec} in dithering mode with an integration time per pointing of 90\,{\it sec}, and Table\,\ref{coords_table} lists the total integration times. More details of the NACO observations, as well as data reduction, are given in \citet{2007ApJ...659L...9M} and \citet{2008MNRAS.384..886V} for the two galaxies, respectively.

\subsection{ALTAIR/NIRI data}

The other 8 targets come from a recent multi-epoch survey using Gemini-North during 2008-2012. The pixel scale is 0.022\textacutedbl$\rm pix^{-1}$, yielding a FOV of 22\textacutedbl. These data also have a final resolution of $\sim$\,0.1\textacutedbl. Each individual frame has an exposure time of $\sim$\,30\,{\it sec}, and in this case separate sky-frames were taken for sky subtraction in addition to on-target dithering. Refer to Table\,\ref{coords_table} for details of observations.

Our Gemini data were reduced using {\tt IRAF}\footnote{ {\tt IRAF} is distributed by the National Optical Astronomy Observatories, which are operated by the Association of Universities for Research in Astronomy, Inc., under cooperative agreement with the National Science Foundation.}-based tasks including flat-fielding and sky subtraction. Individual frames with significantly lower quality PSFs were excluded, as well as some with abnormal electronic noise.  A weaker, horizontal stripe pattern was still evident and was removed by a custom-made de-striping algorithm. The final images were produced by co-adding  by average-combining the individual frames from different observing runs after shifting them to a common reference.  Table\,\ref{coords_table} again lists the effective integration times per target. The astrometry calibration was performed with the {\tt IRAF} task {\tt CCMAP}:  we downloaded archival $HST$/ACS data of the fields and accurately re-calibrated these using the Guide Star Catalog II, and then added the World Coordinate System (WCS) into the FITS headers of our NIR images using the larger FOV $HST$ frames as reference images. 

\subsection{NOTCam data}

We obtained $K$-band images of all the Gemini targets using the Nordic Optical Telescope (NOT) NIR Camera and spectrograph (NOTCam). In the wide field imaging mode, NOTCam has a pixel scale of 0.234\textacutedbl$\rm pix^{-1}$ and a FOV of 4'. The images were flat-field corrected, sky subtracted and combined using {\tt IRAF}-based tasks. The NOTCam images were used as intermediate images to determine photometric zero-points for the smaller FOV NIRI images.

\section{The SSC candidates}\label{analysis}

\subsection{Object detection and photometry}\label{detect}

For object detection we ran {\tt SExtractor v2.5.0} \citep{1996A&AS..117..393B} on unsharp-masked versions of the images  (see the top-right panel of Figure\,\ref{SSC_18} in the case of IRAS\,18293-3413); unsharp masking was done in order to make point-source detection more uniform in varying background conditions,  though we stress that photometry (see below) was done always on original images. Critical parameters for detection were tuned to minimise spurious sources and to include faint and also extended sources in the output catalogues. A threshold of $\sim$\,1.5$\sigma$ above the background RMS noise combined with an upper value of the minimum number of $\sim$~10 adjacent pixels above threshold were eventually chosen.

We performed aperture photometry on the combined images using the task {\tt IRAF/PHOT}  with 3 and 5 pixel aperture radii (approximately equal to a FWHM $\sim$\,0.1\textacutedbl\,of a point source) and  sky annuli from 5 to 7 and 7 to 10 pixels ($\sim$\,0.06\textacutedbl\,wide) for the NACO and the ALTAIR/NIRI data, respectively. A small sky annulus is necessary for a good sky sampling in the strongly varying background in between larger-scale features of the galaxy, while small apertures are needed to minimise blending effects, especially in the case of the crowded SSC populations detected in the NACO data of IRAS\,18293-3413 (hence the smaller apertures with NACO data).  

Since the apertures are small, aperture corrections are essential.  In addition, the PSF shape is expected to vary across the frames as a function of distance from the star used as the AO-reference. Aperture corrections were determined based on the curve-of-growth (out to 1.0") of sufficiently bright and isolated stars at various locations throughout the images. In the case of NACO data the aperture corrections were found to be dependent on the distance of the AO reference star, while in the Gemini data, where the laser guide star is located in the centre of the field within the target galaxy, no such systematics were found.  These aperture corrections  $a_c$, as a function of radial position from the AO star if required, were then applied to measurements of candidate clusters using an aperture of 0.1" radius.   The NACO data $a_c$ values applied to individual candidate SSCs range from $-1.23$ to $-2.07$ mag and for the ALTAIR/NIRI data we adopted an averaged constant $a_c = - 1.23$ mag. We estimate the uncertainty of the aperture correction in a single frame to be typically $\sim$\,0.3 mag.  Note that these aperture correction methods implicitly assume that we are generally detecting point sources, which is supported by our simulations described in Sections\,\ref{selection},\,\ref{completeness} and\,\ref{blending_sec}. 

\begin{figure*}
  \begin{tabular}{c}
  \resizebox{0.42\hsize}{!}{\rotatebox{0}{\includegraphics{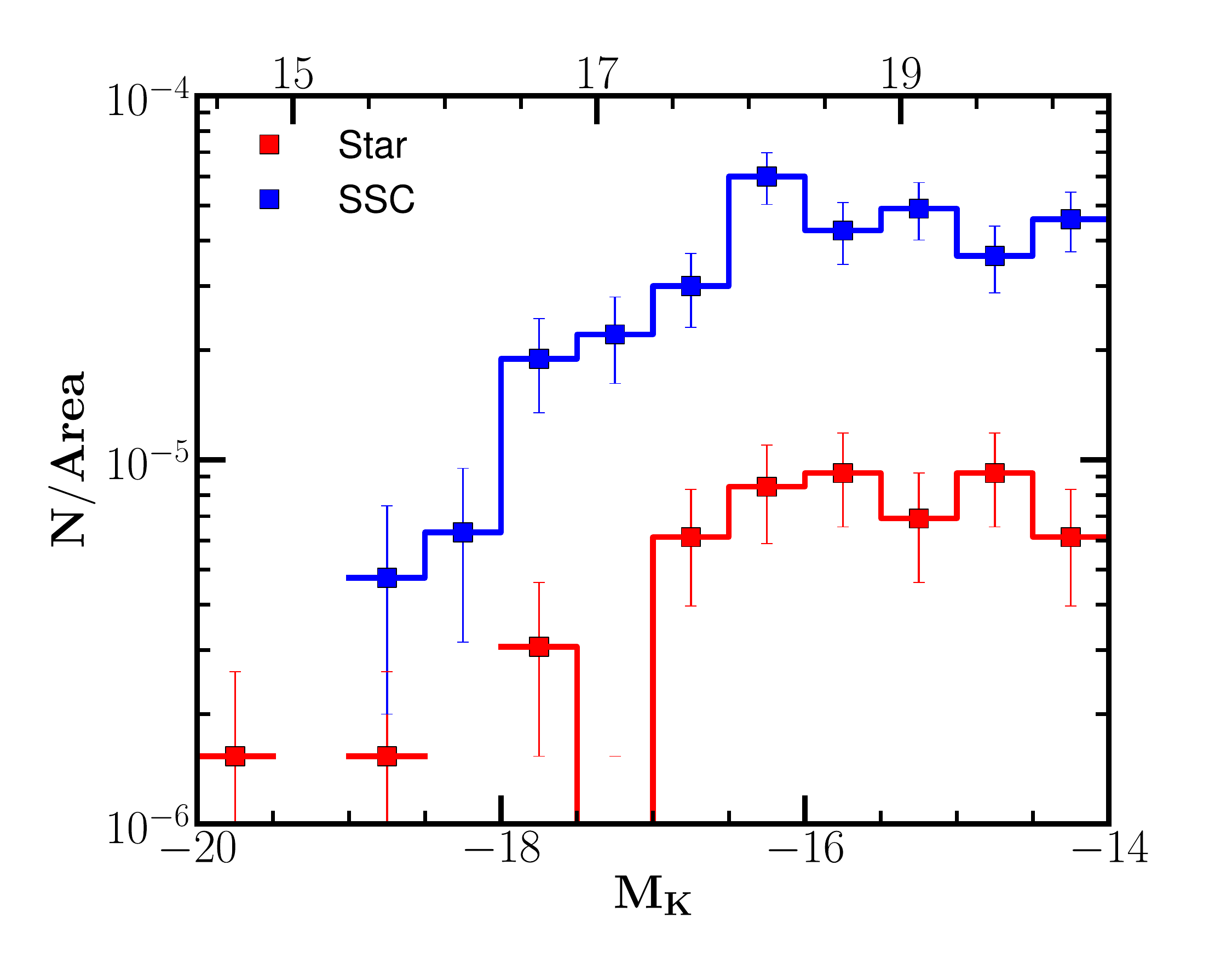}}}
  \resizebox{0.42\hsize}{!}{\rotatebox{0}{\includegraphics{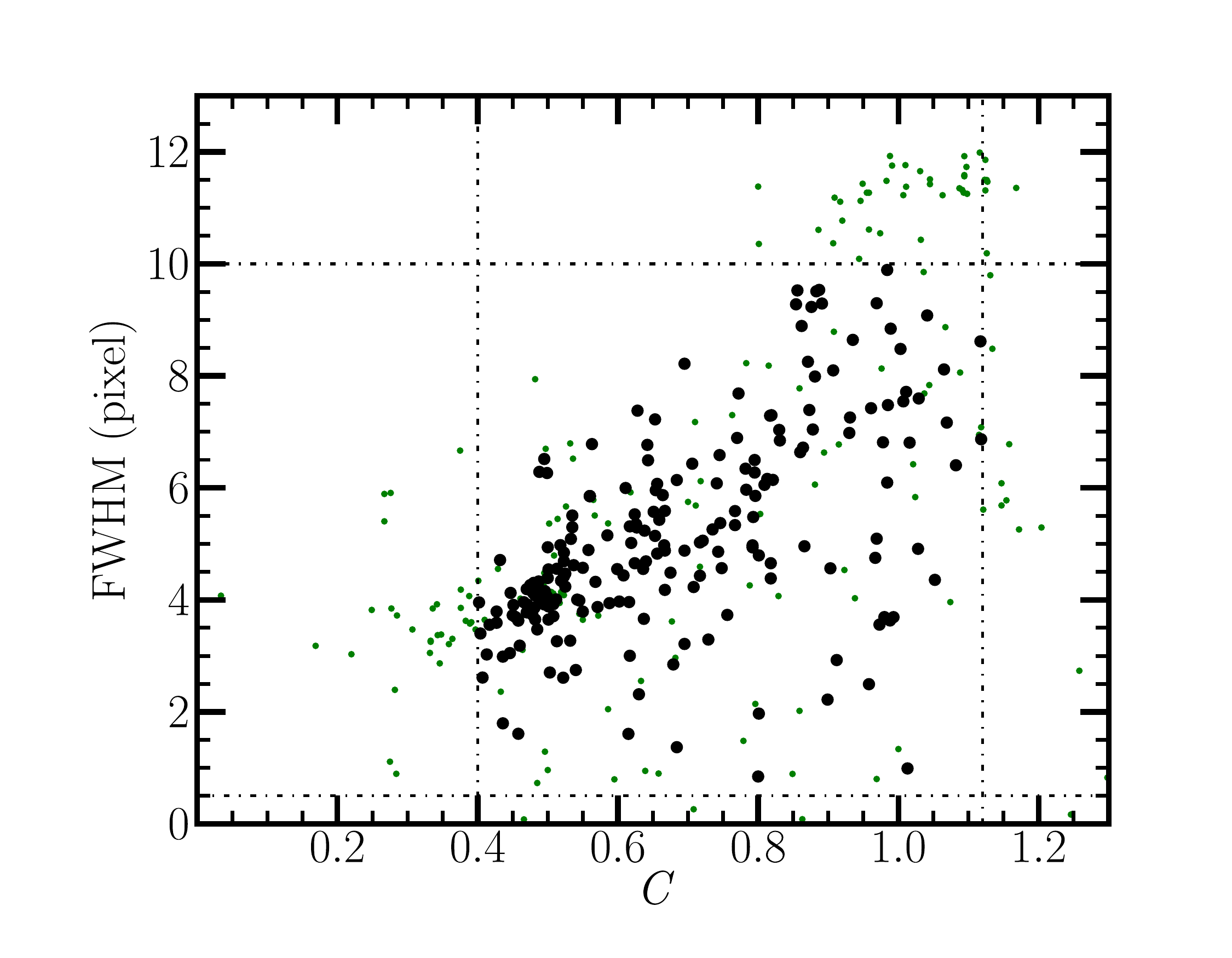}}}\\
 \end{tabular}
\caption{\small {\em Left:}  Raw surface densities of detected objects within the galaxy area of frame IRAS\,18293-3413 are shown in blue, while those of objects found {\em outside} the galaxy area in the same frame are shown in red.  As the latter are likely foreground stars, it is seen that stellar foreground contamination is insignificant even in this low galactic latitude case.  Apparent $K$-band magnitude scale is shown at the top. {\em Right:} An example of SSC candidate selection for the same field of IRAS\,18293-3413.  A FHWM vs concentration index {\it C} plot of the detected objects from {\tt SExtractor} with the cutoffs for candidate  selection indicated.  The limits are checked with simulations to be consistent with the parameter space of extracted {\em intrinsic} point sources ($\sim$\,4.5\,pix in this case) in the fields. The filled black points represent the final SSC candidates, whereas all detected objects are shown in green.}
\label{select_plot}
\end{figure*}

Finally, photometric zero-points were referred to the 2MASS $K_S$-band point source catalogue using 2MASS common stars in the FOV, making our magnitude system the same as that of the 2MASS Vega-based $K_{S}$\,filter. In cases where the 2MASS point-sources were outside the small FOV of the Gemini data, NOTCam $K$-band images were used as intermediate images to estimate the zero-points.  For MCG\,+08-11-002 and CGCG\,049-057, comparing the total integrated flux of the galaxy with the corresponding 2MASS magnitude was the only option to estimate their zero-points. We estimate the absolute calibration to be accurate to $\sim$\,0.4 mag including 2MASS catalogue photometric uncertainties and systematic errors in our photometry and aperture corrections.

\subsection{Cluster candidate selection}
\label{selection}

\begin{table}
\centering
\begin{minipage}{100mm}
\begin{tabular}{|lc|c|c}\\
\hline
\\
\multicolumn{1}{|c}{Name} &
\multicolumn{1}{|c}{N(SSC)} &
\multicolumn{2}{|c}{Comp.limit ($K$-band)} \\
\cr &($\sigma_{m}$~$\leqslant$~0.25)&App mag&Abs mag\\\hline
\hline
\cr IC 694&81& 19.7& -13.6\
\cr NGC 3690&81&19.4&-13.9\
\cr CGCG 049-057&10&19.4&-14.4\
\cr IRAS F17578-0400&45&20.4&-13.4\
\cr IRAS F17138-1017&60&20.0&-14.3\
\cr IRAS 18293-3413&204&19.9&-14.5\
\cr MCG~+08-11-002&12&19.2&-15.3\
\cr IRAS F16516-0948&41&19.9&-14.9\
\cr IC 883& 29&20.2&-14.8\
\cr IRAS 19115-2124&56&20.1&-16.5\
\cr &&\\
\hline
\end{tabular}
\end{minipage}
\caption{\small After imposing our selection criteria,  the final number of SSC candidates for each target is given. We tabulate also the apparent and absolute magnitudes of the 50\,\% completeness level corresponding to the background region where more than $\approx$\,50\,\% of the data points are below the contour level limiting that region (see text).  The brighter absolute magnitude completeness limit for IRAS\,19115-2124 is due to its much larger distance compared to the other targets.}
\label{comp_lim}
\end{table}

Deciding whether a detected source is a SSC candidate or not is challenging when working with a single filter  and dealing with a sample of distant targets where both stars and star clusters have similar PSF sizes. In our case, the following steps were carried out to generate the final $K$-band SSC catalogue for each target.

\begin{enumerate}

\item First of all, only objects falling on detectable optical or NIR emission from the galaxy within the frames were considered.  The edges of frames where the noise is higher were also excluded in case the host galaxy extended there.  After this process, we assumed that the catalogue is mainly composed of real sources since we had already tuned the detection parameters to avoid false detections in the inner regions of the frames.

\item Foreground contamination by Milky Way stars is a real possibility, especially  for targets with low galactic latitude ({\it b}\,$<$\,$|20^{\circ}|$, see Table\,\ref{coords_table}).  Therefore, we estimated the  potential effect of  contamination in our data by using the Besan\c{c}on model  \citep{2003A&A...409..523R}.   IRAS\,19115-2124, IRAS\,18293-3413, and IRAS\,17578-0400 yield the largest numbers of expected $K$-band stars according to the model, but even in these fields the maximum numbers of contaminating stars at $K$\,$\sim$\,16 mag corresponding to the brightest magnitude bins of SSCs, are at most at $\sim 10$\,\% of the total SSC candidate numbers within a typical 20\textacutedbl\,$\times$\,20\textacutedbl area of the galaxies. At fainter magnitudes star counts are at negligible levels, as are counts at all magnitude bins for the rest of the fields, according to the models. However, as a consistency check, we also counted obvious point sources in the sky areas {\em outside} of the galaxies in our fields, derived their surface densities in magnitude bins, and compared to object numbers within the host galaxy area.   As an example we show the low galactic latitude IRAS\,18293-3413 case in the left panel of Fig.~\ref{select_plot}. These numbers are consistent with the Besan\c{c}on model predictions at bright magnitudes, but tend to be somewhat larger at $K>16$: it could, for example, be that there are some real SSCs well outside of the projected galaxy area of our targets, or that the models do not probe Milky Way stars at faint enough levels.  Though there might be some contamination at the brightest $K\sim 15-16$ bins, these would be "obvious stars" that are removed in the final selection step (v) below  and at fainter levels the star numbers still are not large enough to make any significant difference to the ultimately derived luminosity functions of SSCs.  We hence conclude that foreground contamination is not significant for our results, and no further star vs. SSC separation was attempted for our dataset.  

\item Photometric uncertainties were taken into account fairly conservatively to have a robust list of SSC candidates:  we excluded all detections having errors $\sigma_{m}$\,$>$\,0.25 mag.  We note that excluding less secure detections can in principle bias eventual LF values since fainter objects will not be included.  However,  we checked that if we had selected $\sigma_{m}$\,$<$\,1.0\,mag sources instead, we would have had only between 6 to 14\,\% more SSC candidates, and the final LF shapes and slope changes would have been negligible within the completeness levels considered in the analysis.

\item We excluded objects that appeared to be extended in order to remove background galaxies and resolved HII regions, while also taking into account the fact that fainter point sources will appear to have a broader FWHM when extracted on top of  the varying diffuse background of the galaxy.   For a candidate to be included in our photometric catalogue, we imposed cuts on the values of a "concentration index" {\it C}, defined as the difference in magnitudes between the 3 and 5 pixel radius apertures ($m_{3px} - m_{5px}$), as well as cuts on the FWHM.  A small value of the index ({\it C}~$\sim$~0.5) typically results from a high S/N point source, such as a foreground star, whereas larger values ({\it C}~$\sim$~1.0 and above) indicate potentially extended objects.  Since the value of $C$  is less accurate for fainter sources, we also used the FWHM of the detection, as measured by the task {\tt RADPROF} in pixels to make the selections more robust.  Plots such as that shown in  the right panel of Figure\,\ref{select_plot}, were generated for each target field to help us distinguish between truly fuzzy objects and unresolved SSC candidates at different S/N levels. 

The cutoff values were decided with the help of simulations and also by-eye checks interactively for consistency.  As part of Monte Carlo simulations to define completeness corrections (described in Section~\ref{completeness}) simulated intrinsic point source PSFs were extracted from the real data frames, and the cutoffs in FWHM and $C$ were adjusted to encompass the {\em output parameter space} of the detections of the {\em input point sources} in the simulations made on the real data frames.
The differing AO corrections on different datasets introduces some variation, but  in most of the cases SSC candidate selection included objects having values of $0.5\lesssim{\it C}\lesssim1.0$  
and $0.6\lesssim \rm FWHM \lesssim10.2$ pixels.   In the case of IRAS\,F16516-0948 all objects are generally elongated, possibly due to a degraded AO-correction; however, since all analysis was done using PSFs constructed from bright stars on the frame itself, this does not bias the selection in any way.

Note that retaining only unresolved point sources at this step does not exclude the possibility of blending  discrete SSCs smaller than our resolution PSF into one detected SSC candidate -- this is discussed in more detail in Section~\ref{blending_sec}. 

\item Finally, we made a visual inspection of the images with the selected candidates in order to remove the nucleus of the galaxy from the final catalogue and to remove several obvious very bright stars which happened to fall on top of the galaxies.

\end{enumerate}

After applying the selection criteria detailed above we obtained the final catalogue for each target field.  The total number of SSC candidates per LIRG are listed in the second column of Table\,\ref{comp_lim}.  Figure\,\ref{SSC_18}  shows, in the lower panels, the spatial distribution of the SSC candidates in the case of IRAS\,18293-3413:  all the sources detected originally by {\tt SExtractor} are marked on the image, but only those which met all the selection criteria above are in red (most are excluded because of their magnitude uncertainty in this case).

\subsection{Completeness correction}\label{completeness}

\begin{figure}
\resizebox{1.\hsize}{!}{\rotatebox{0}{\includegraphics{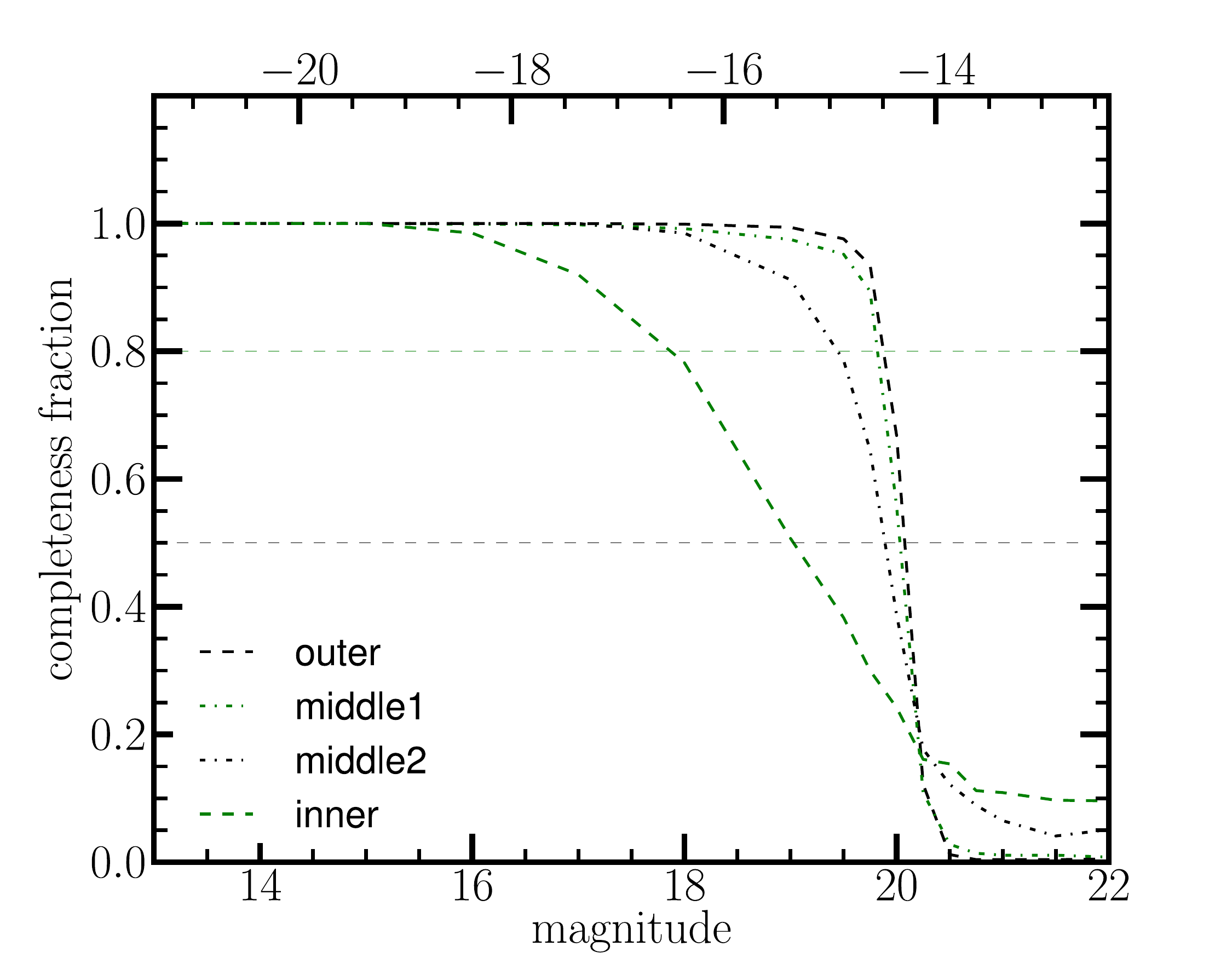}}}\\
\resizebox{1.\hsize}{!}{\rotatebox{0}{\includegraphics{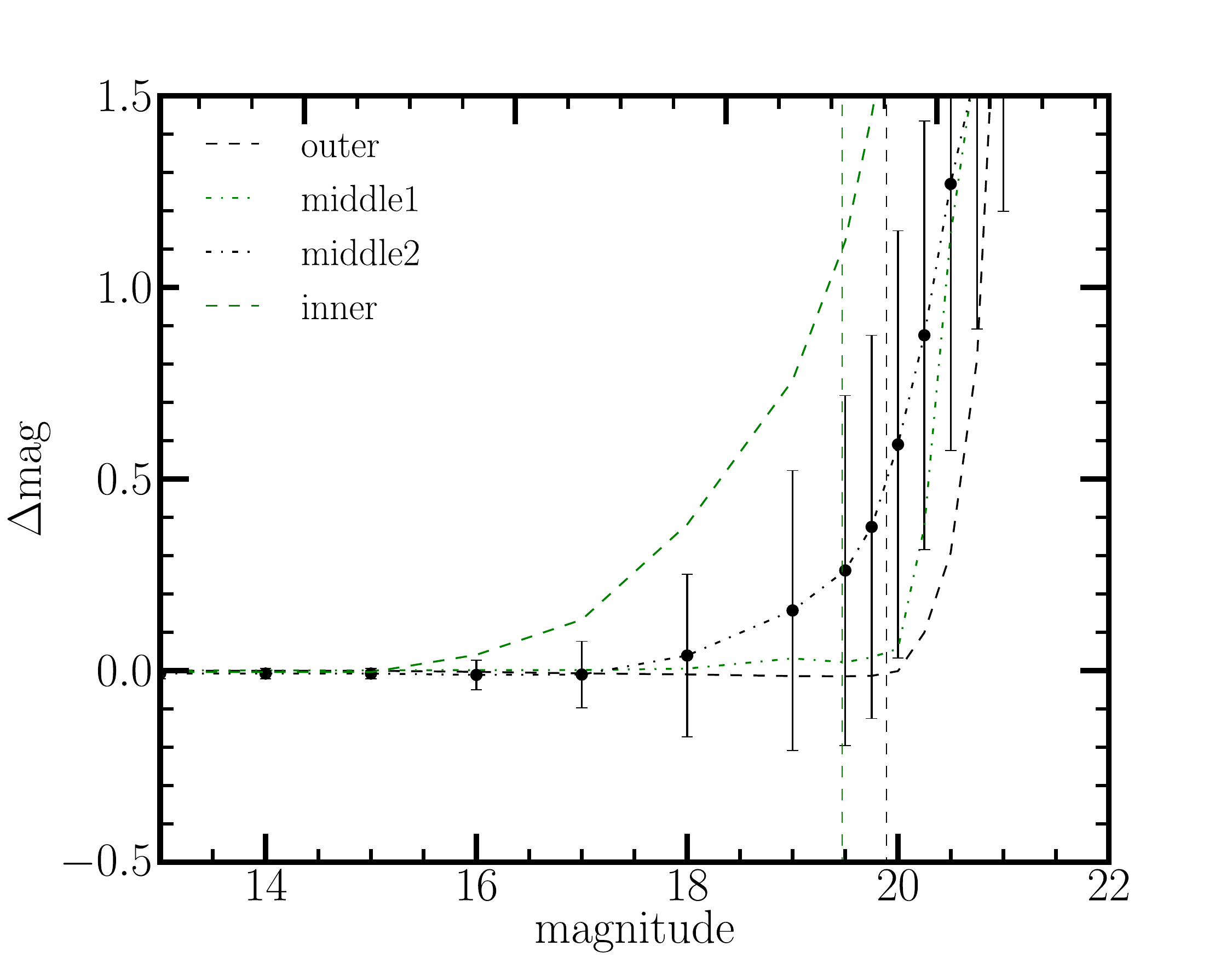}}}
\caption{\small The results of Monte Carlo completeness simulations for IRAS\,18293-3413 within regions of different background levels. The green dashed line corresponds to the innermost region with the highest background.  {\it Upper panel:}  The fraction of simulated point sources recovered as SSC candidates in the simulation as a function of apparent and absolute $K$ magnitude in the lower and upper axes, respectively.  
The 50\,\% and 80\,\% completeness limits corresponding to the \textgravedbl middle2\textacutedbl\ region are a reasonable approximation of overall completeness limits of point sources in IRAS\,18293-3413; they are shown as the horizontal dashed lines.
{\it Lower panel:} The y-axis plots the input minus output magnitude as a function of the input magnitude in the same MC simulation. The error bars reflect the scatter of this difference in the simulation and is plotted for one curve only for clarity. The vertical dashed lines show the two completeness levels determined above from the \textgravedbl middle2\textacutedbl\, region.}
\label{inc_plot}
\end{figure}

The raw SSC catalogues obtained in the previous section are affected by incompleteness due to photometric detection limits of the observations; the varying and complicated background resulting from the different diffuse components of the host galaxies;  and potential crowding of sources. To correct the data for incompleteness bias, we ran Monte Carlo (MC) completeness simulations with each science image. A PSF model for each target field was first created.  The LIRGs MCG+08-11-002 and CGCG\,049-057 did not have isolated stars in their fields, and for them we used a representative PSF model from other fields having a similar distance from their tip-tilt reference star, and hence similar AO-correction based from the reference star.  Using the model, we created artificial stars with {\tt IRAF/DAOPHOT} which were used in the simulation. 

The main idea of the simulation is to record how many of the input objects, added randomly to the science images, are detected by {\tt SExtractor} at a given magnitude range with exactly the same manner and configuration parameters as the ones chosen for optimal detection in $\S$\,\ref{detect}, as well as with identical selection criteria to be included as an SSC candidate.  As mentioned above, these simulations were also used to define the latter selection criteria since the input source is by definition a point-source.   We also checked the simulations for varying PSF sizes and verified that PSFs in a realistic range of point source sizes in the target fields do not result in corrections which would change any of the main results concerning the shapes of the eventual LFs.  

The simulation ran from 13 to 22 $K$-mag in steps of 0.5 mag. We generated 1000 random positions of artificial star centroids. To avoid systematic errors, we used a new subset of random positions for each magnitude step.  In addition we ran the simulation separately at different background levels of the field.  These levels were determined  by defining three (four in cases of IRAS\,19115-2124 and  IRAS\,18293-3413, the latter having the largest dynamical range between the background and point source brightness) approximately equal ranges in the pixel values of a smoothed background map in a logarithmic scale, ranging from an essentially empty sky to the LIRG's core, but excluding the nucleus itself.   Figure\,\ref{inc_plot}, upper panel, shows as an example the resulting completeness curves for the case of IRAS\,18293-3413. 
The simulations were also used to test the accuracy of the photometry for systematic effects.  The lower panel of Fig.\,\ref{inc_plot} shows the difference of input and output magnitudes of the detected sources.  In the fainter bins, magnitudes cannot be measured reliably anymore due to very variable background, crowding, and inaccurate aperture corrections.  However, the $\Delta \rm mag$ values become larger than our typical overall photometric uncertainties only at or below the completeness limits relevant to each galaxy and background region used for subsequent LF analysis, and any corrections for these possible systematics are not attempted.

Each individual SSC photometric data point was then corrected for the incompleteness bias with respect to its observed magnitude as well as its location in the complex background field, as long as its magnitude value was above the 50\,\% completeness limit of the region in question.  Representative completeness values are shown in Table\,\ref{comp_lim}. After this correction, we are ready to construct the luminosity functions of the SSCs in each target field. 

\begin{figure*}
  \centering
  \begin{tabular}{cc}
\resizebox{0.40\hsize}{!}{\rotatebox{0}{\includegraphics{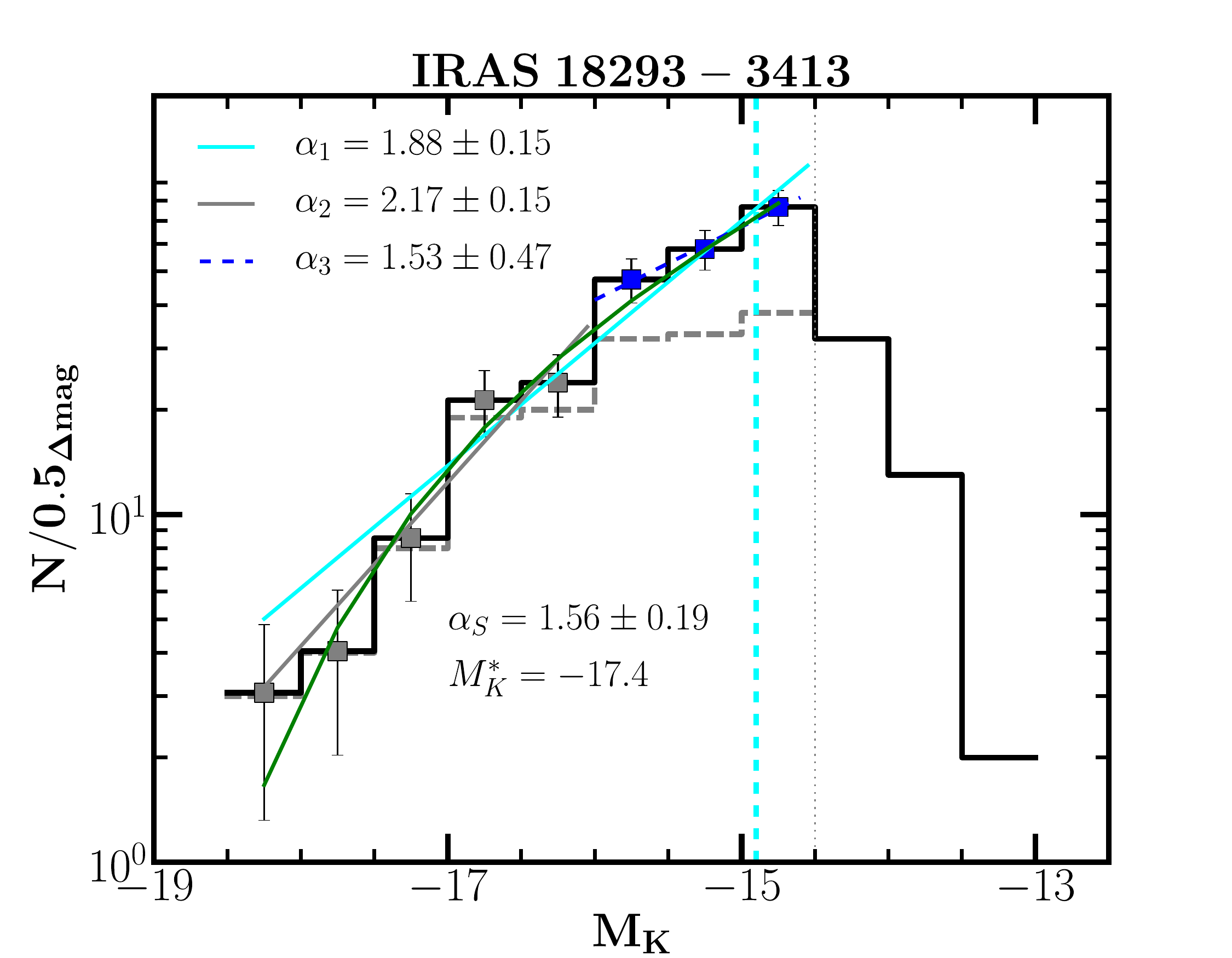}}}
\resizebox{0.40\hsize}{!}{\rotatebox{0}{\includegraphics{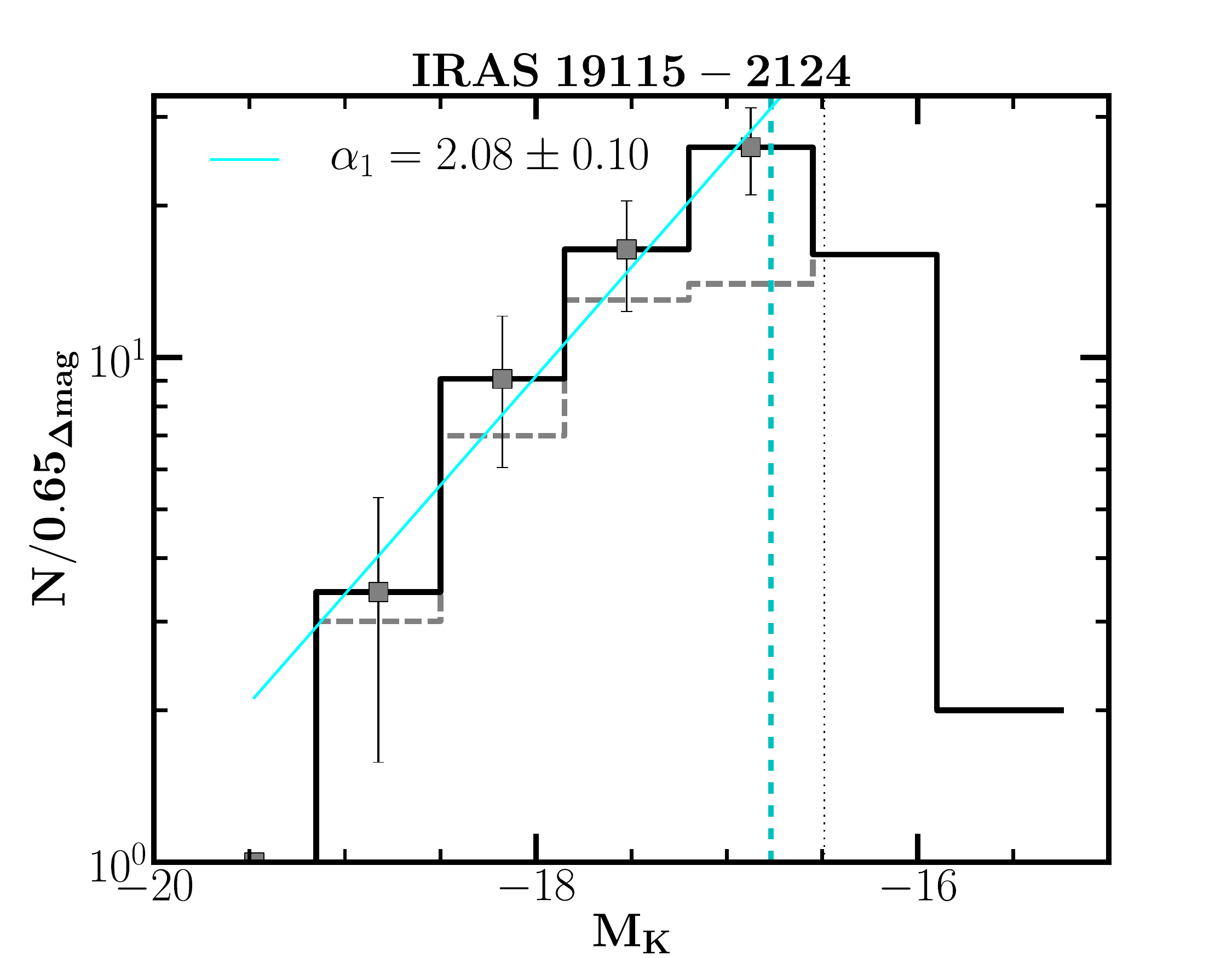}}}\\
\resizebox{0.40\hsize}{!}{\rotatebox{0}{\includegraphics{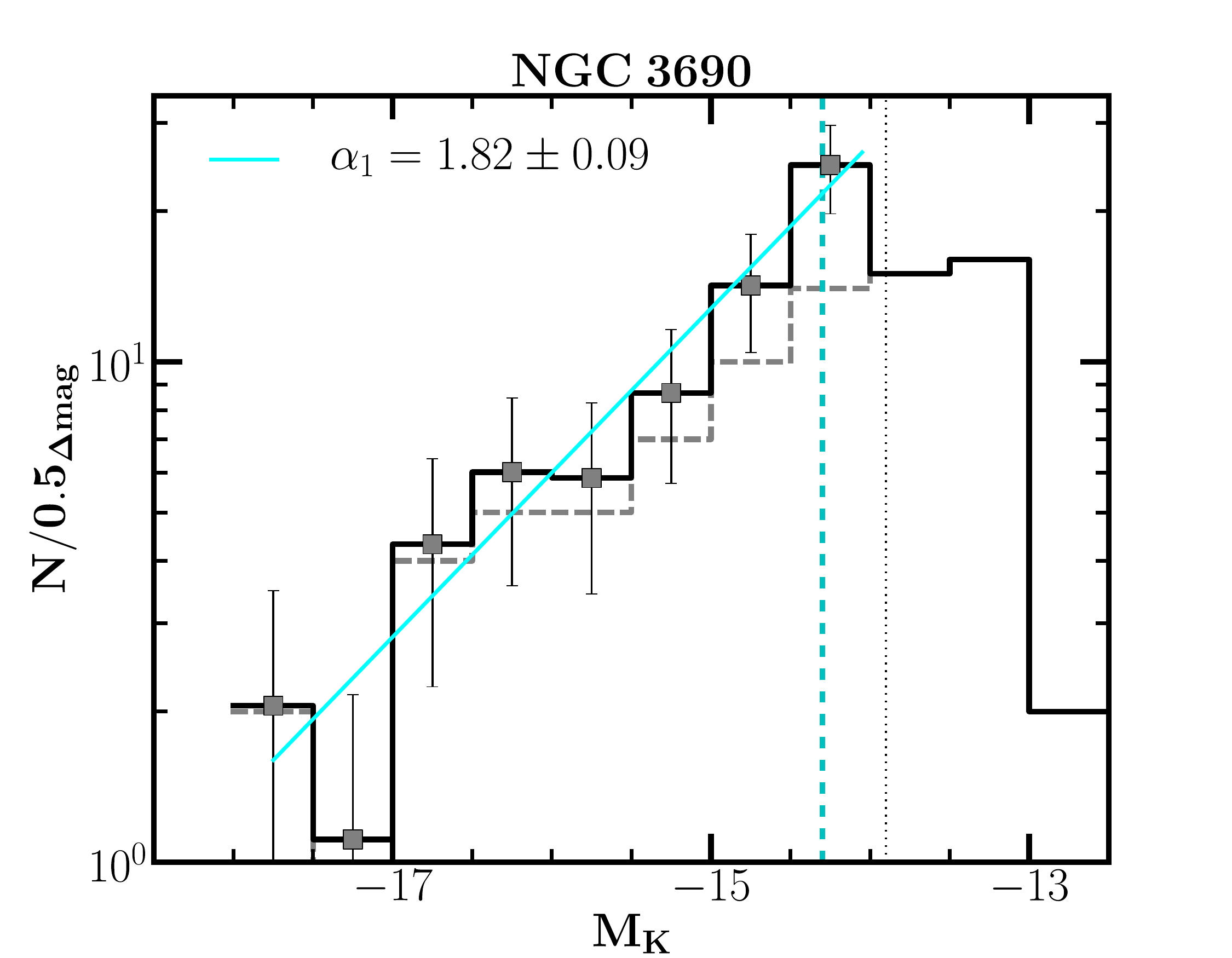}}}
\resizebox{0.40\hsize}{!}{\rotatebox{0}{\includegraphics{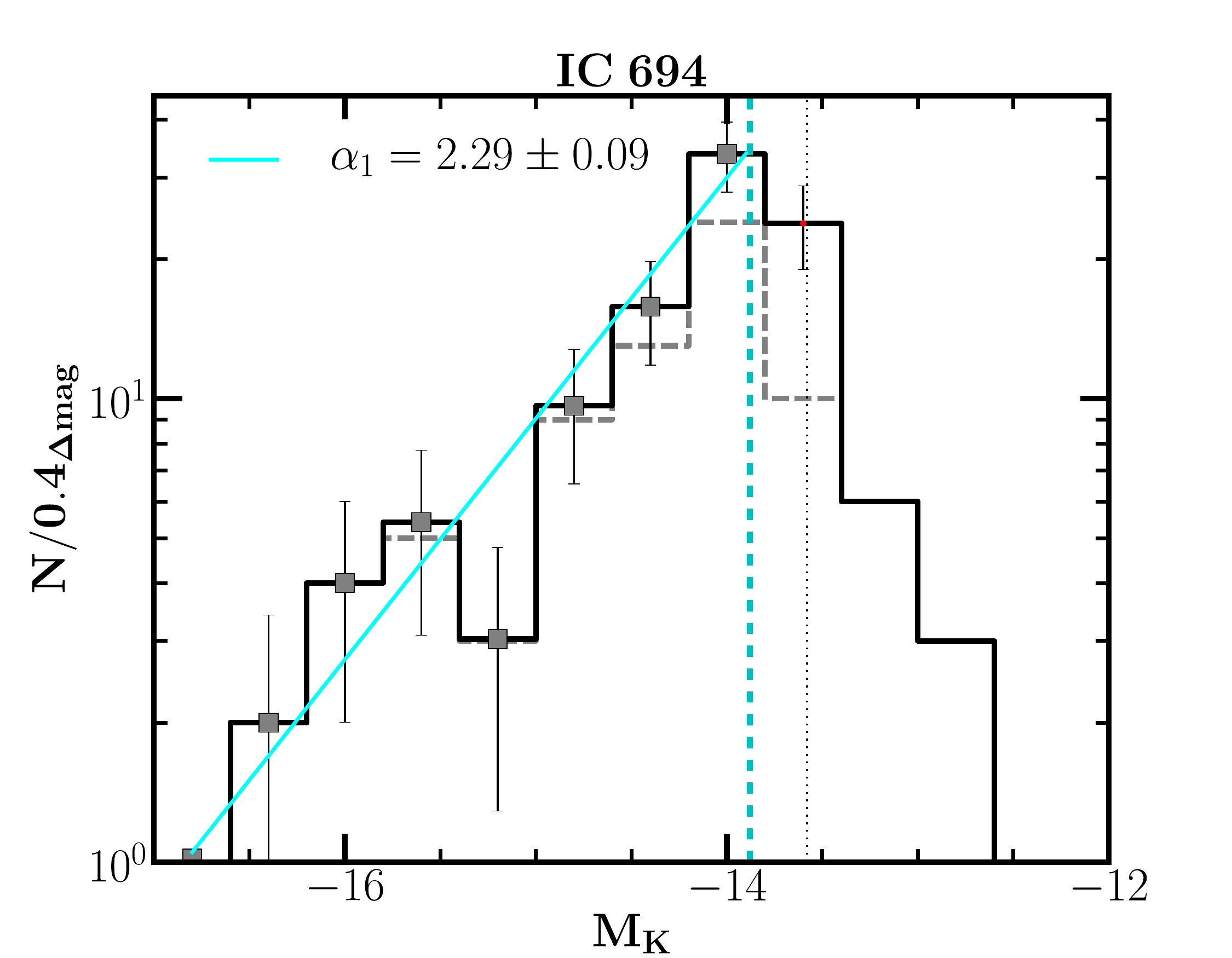}}}\\
\resizebox{0.40\hsize}{!}{\rotatebox{0}{\includegraphics{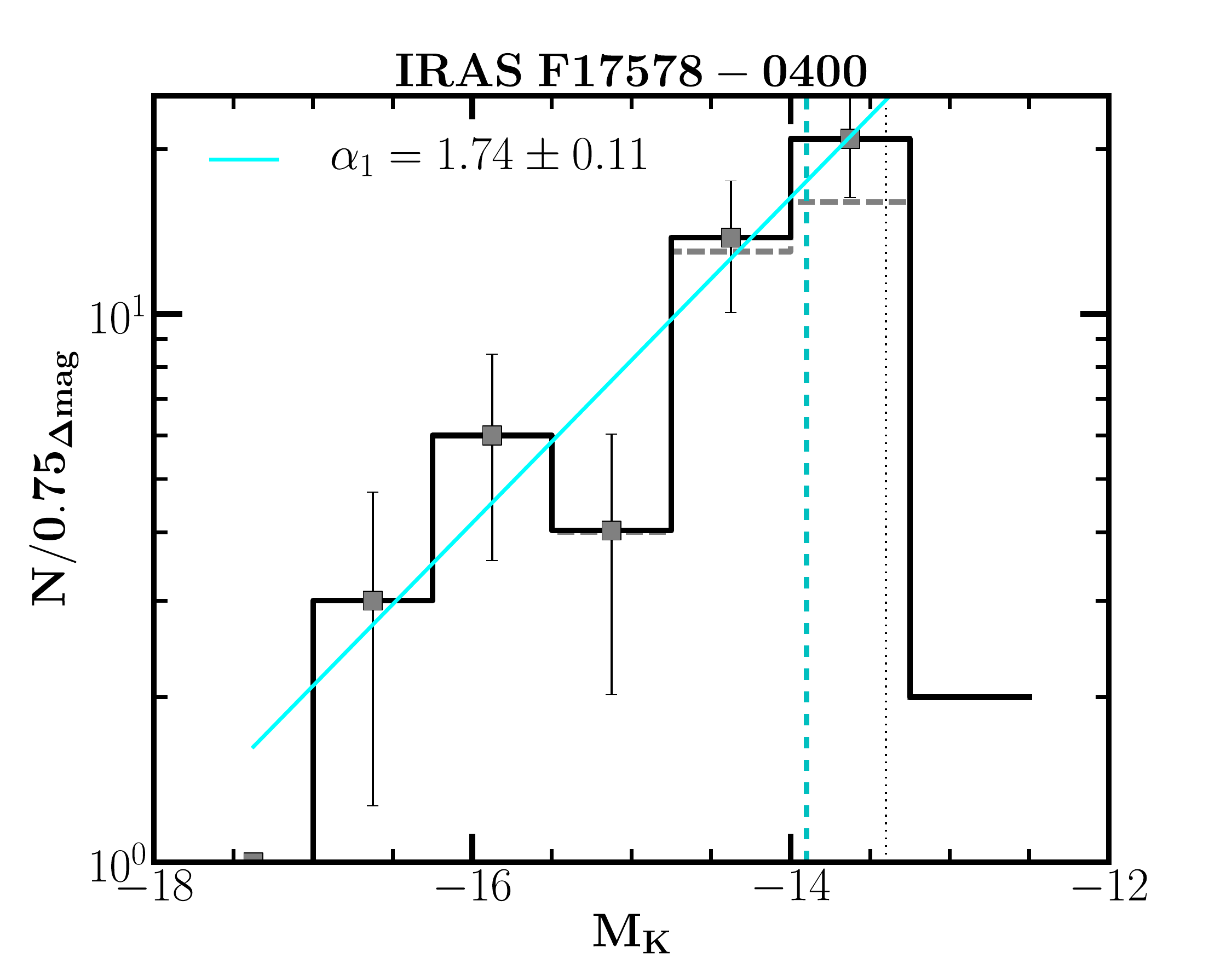}}}
\resizebox{0.40\hsize}{!}{\rotatebox{0}{\includegraphics{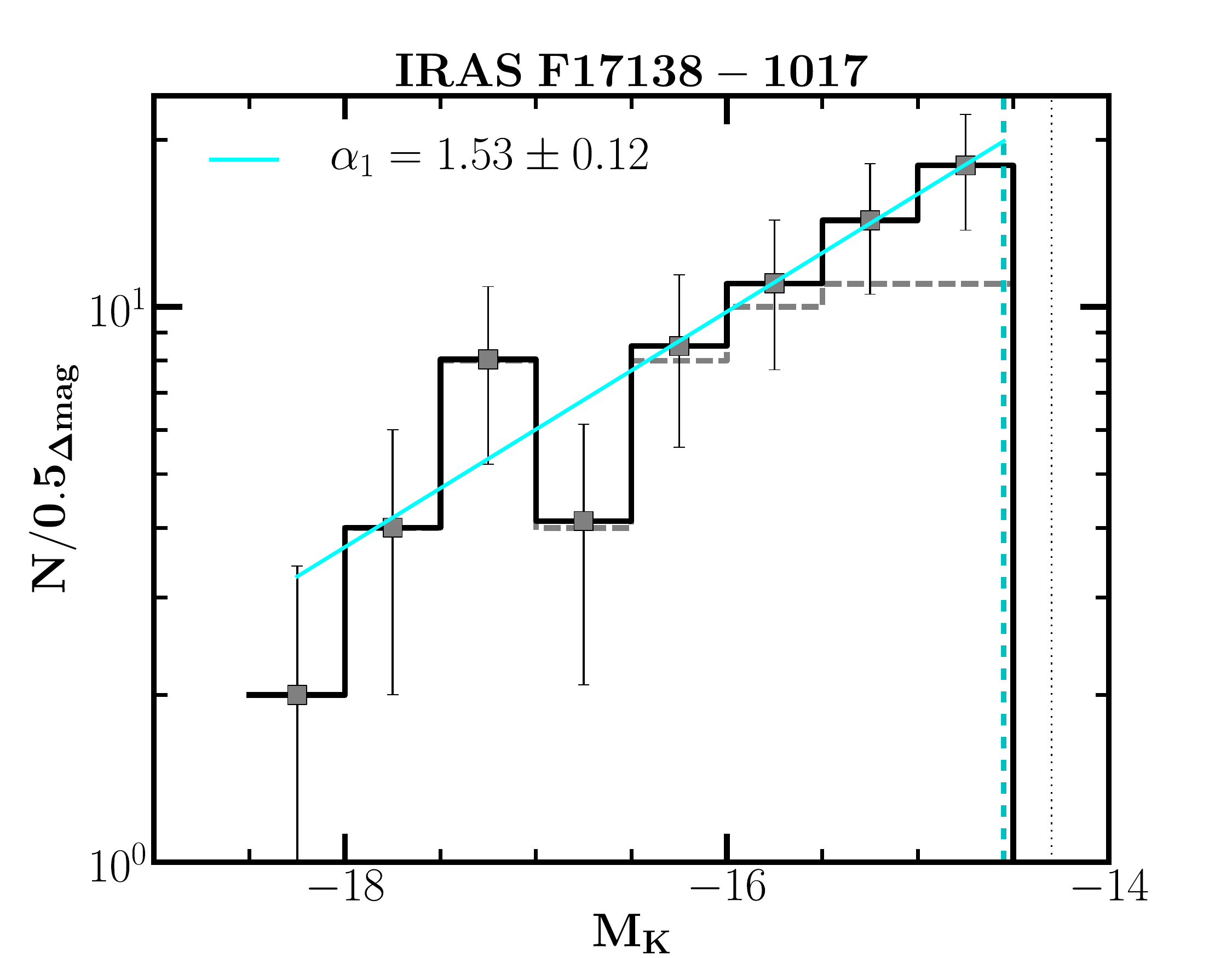}}}\\
\resizebox{0.40\hsize}{!}{\rotatebox{0}{\includegraphics{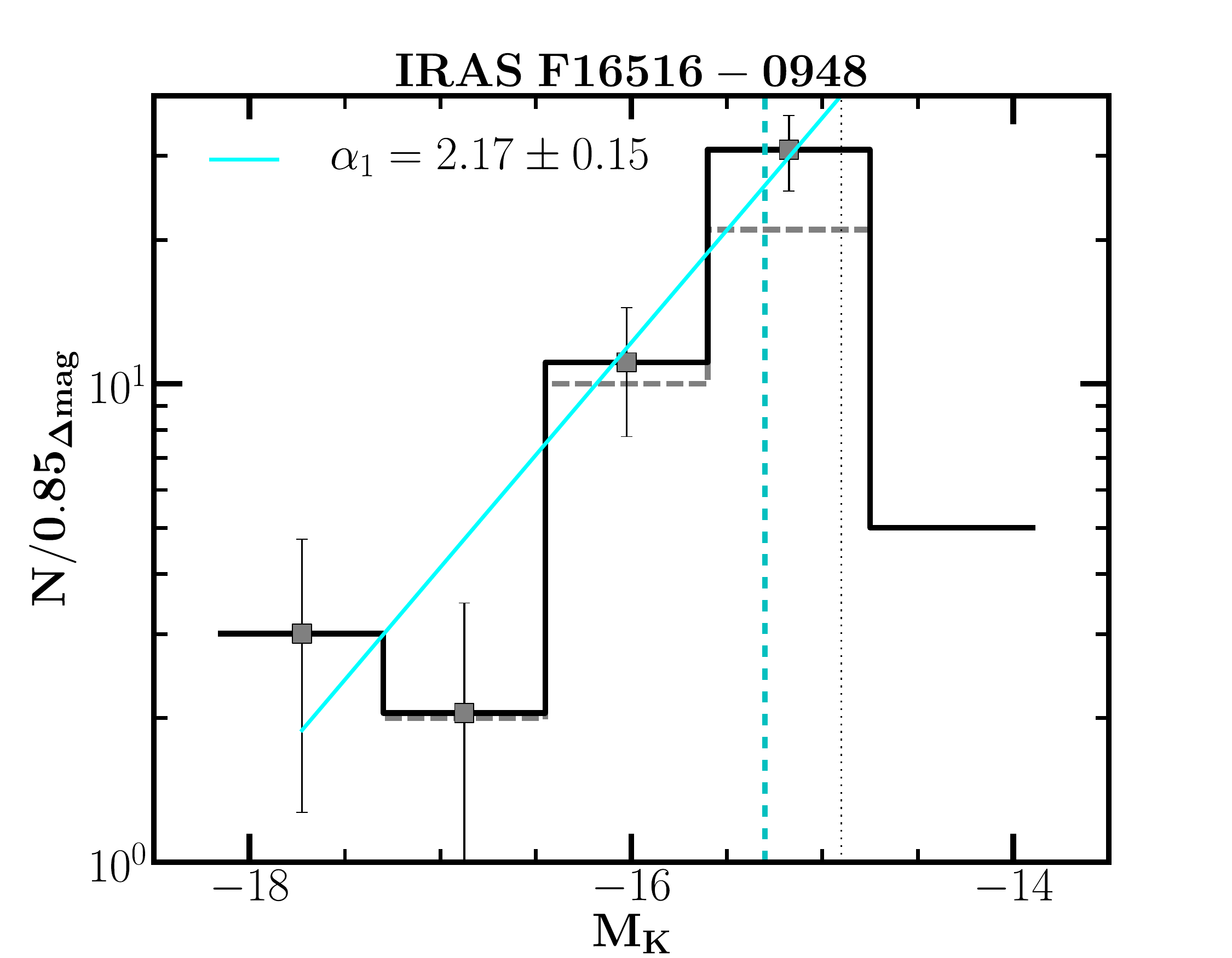}}}
\resizebox{0.40\hsize}{!}{\rotatebox{0}{\includegraphics{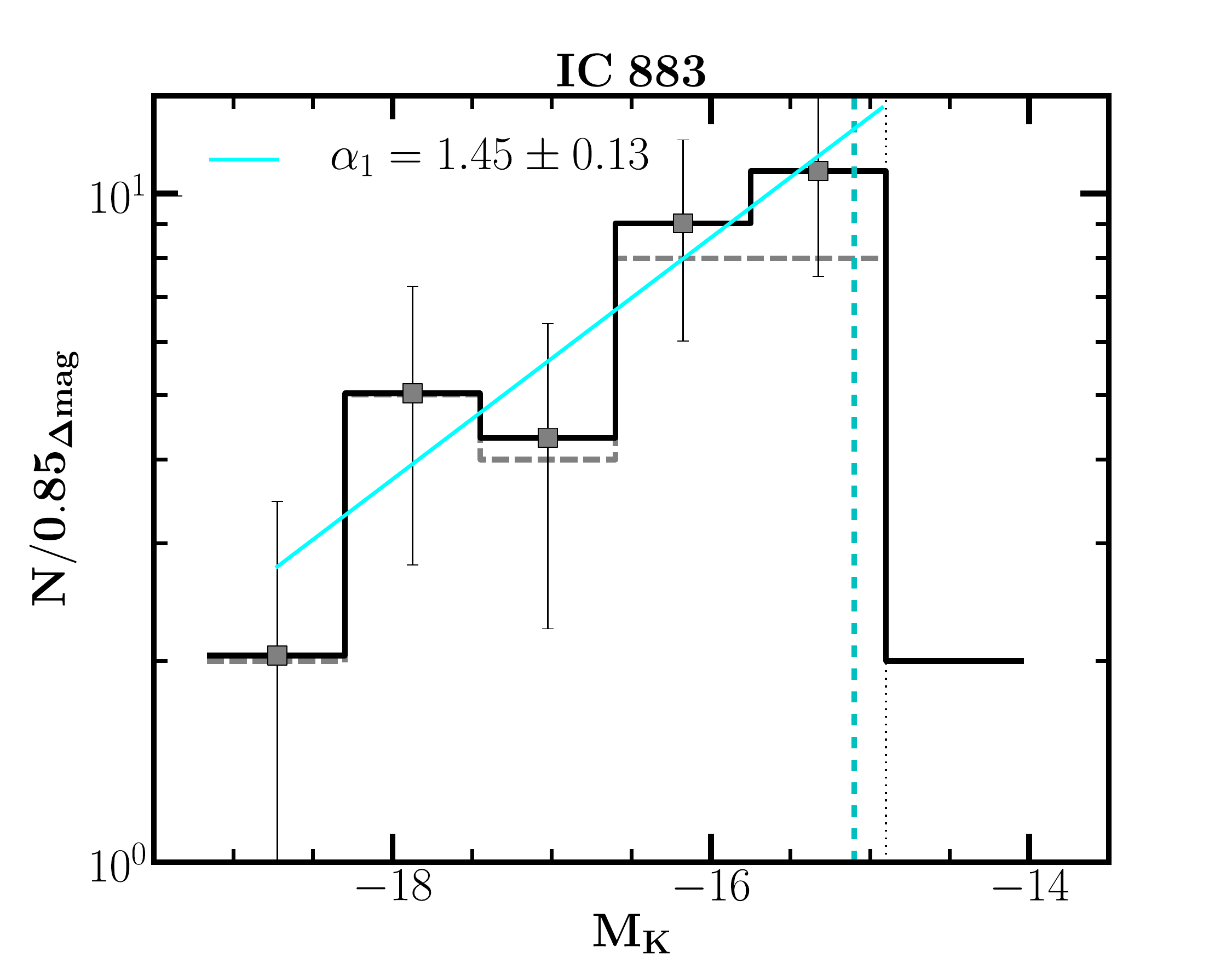}}}\\
 \end{tabular}
\caption{\small $K$-band LFs of the sample using a constant bin size. The black solid line is the incompleteness-corrected LF while the dashed grey line is the original one. The single power-law fit of the data points is represented by the cyan solid line while, in the case of IRAS\,18293-3413, the grey solid one and the dashed blue line result from broken power-law fits,  and the curved green line is a Schechter function fit. The vertical lines mark the 50\,\% (grey) and the 80\,\% (cyan) completeness levels. The y-axes scales are not the same, because the number of SSC candidates differs from one target to another, making the optimal constant bin size different in each case. }

\label{LF_fig}
\end{figure*}

\section{Constructing the $K$-band SSC Luminosity Function}\label{fit_LF}

Once the SSC candidates have been selected for each target, we can construct a binned LF and then fit a function to its shape.  LFs are constructed as a function of absolute  $M_{K}$ magnitude; the distance modulus $m - M$ of the observed sample listed in Table\,\ref{coords_table} was used in the conversion.   Completeness corrections were applied to the counts of SSC candidates as a function of observed magnitude and background region of the detections, as discussed above.   Figure\,\ref{LF_fig} shows all the observed LFs of our targets; both raw and completeness-corrected counts are shown.

After this, we fit a power-law of the form $N(L)dL \sim L^{-\alpha}dL$ in log-log space to the LF shape.  The fitted data points are weighted using their respective Poisson-noise uncertainties $\sqrt{N}$.  In a mag-logN plot, a linear fit is expressed as follows:
\begin{equation}\label{log_log_slope}
 {\rm log}N(M_{K})\,=\,\beta\,M_{K}~+~con
\end{equation}
where the relation between the power-law index $\alpha$ (see Eq.\,\ref{equone}) and the linear slope $\beta$ is \citep[][]{1997ApJ...480..235E}:
\begin{equation} \label{conv_slope}
\alpha\,=\,2.5\beta~+~1
\end{equation}

Due to the very small number of SSC candidates in MCG+08-11-002 and CGCG\,049-057, N=12 and 10 respectively, we did not fit their LFs.

\subsection{Single power-law fits}

\begin{figure}
\resizebox{1.\hsize}{!}{\rotatebox{0}{\includegraphics{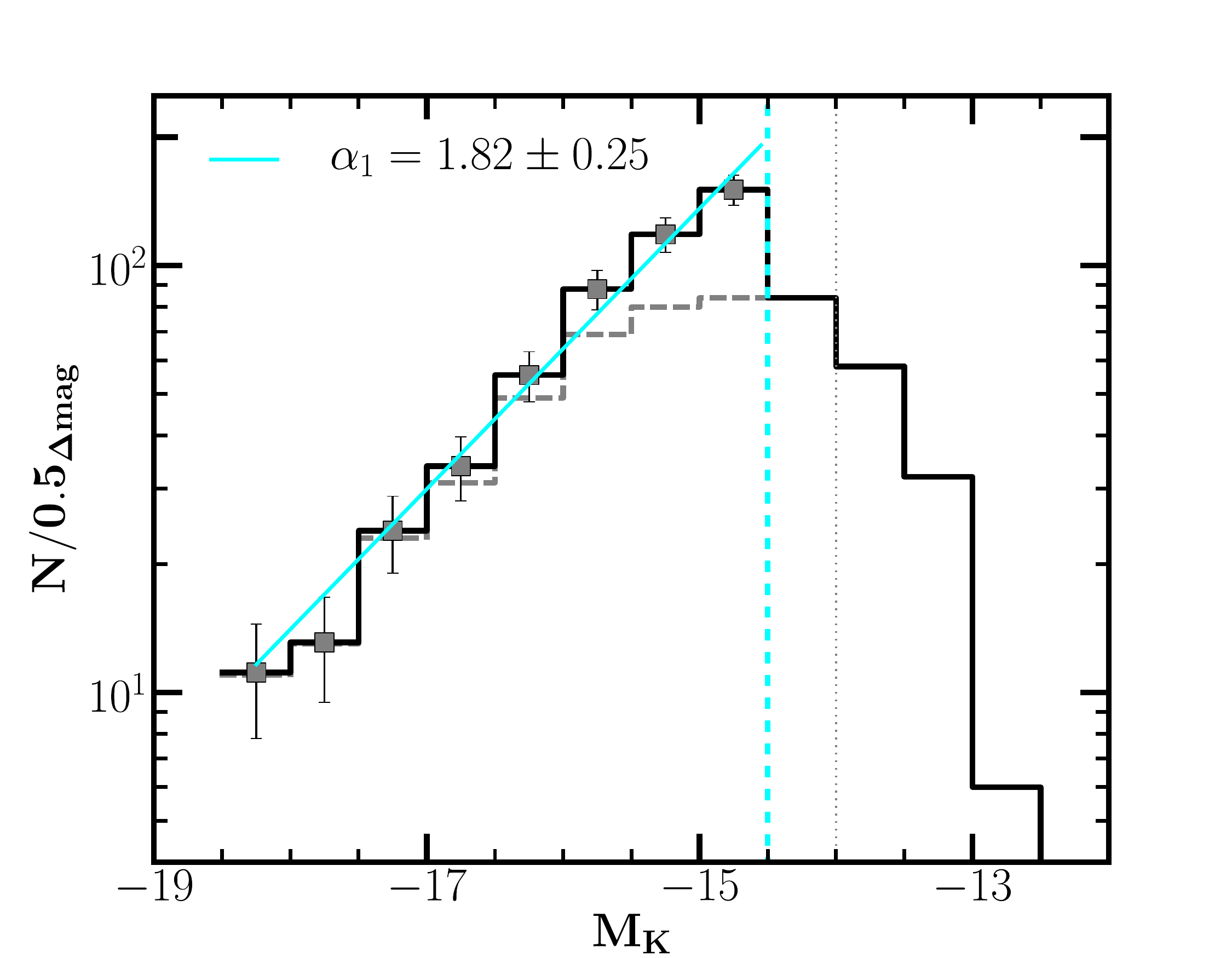}}}\\
\caption{\small $K$-band SSC LF of the sample, except the dataset from IRAS\,19115-2124. The single power-law fit of the data points is represented by the cyan solid line while the vertical lines mark the average values of the 50\,\% (grey) and the 80\,\% (cyan) completeness levels.}
\label{LF_sample}
\end{figure}

As a first step we fitted the SSC candidate LF distributions with a single power-law shape using a constant bin size.  Since the targets do not have the same number of SSC candidates, each galaxy has its own constant bin size to balance having statistically enough sources per bin while trying to maximise the number of bins overall.  
The LF bins were fitted from the brightest bin down to the last bin above the 80\% completeness limit, the data points plotted as squares with error bars in Fig.\,\ref{LF_fig} indicating the bins which were fit.  The resulting power-law indices estimated using Eq.\,\ref{conv_slope} are  shown in Table\,\ref{slope}  as $\alpha_{1}^{con}$, and the fitted LFs are plotted in Fig.\,\ref{LF_fig} as the cyan line. The values of $\alpha_{1}^{con}$ range from 1.5 to 2.4. The average over the sample is  $\alpha_{1}^{con} = 1.87 \pm 0.30$, or $\alpha_{1}^{con} = 1.84 \pm  0.28$ if the most distant LIRG, IRAS\,19115-2124, is excluded. The quoted errors are the formal uncertainty of the fits. Fitting a single power-law function to the LF of the combined dataset down to a $M_K = -14.5$\,mag (which includes the SSCs from all 8 LIRGs in Fig.\,\ref{LF_fig} except for the most distant target IRAS\,19115-2124 which has a significantly different magnitude), gives a slope $\alpha_{1}^{con} = 1.82 \pm 0.25$ (Fig.\,\ref{LF_sample}). This is quite similar to the average slope.

To check whether a single power-law is a good approximation of the LF, we estimated the reduced chi-square statistic $\chi_{red}^{2}$ (the ratio of chi-square $\chi^{2}$ and the degrees of freedom of the dataset) in each case. In most cases $\chi_{red}^{2}$ values (see Table\,\ref{slope}) indicate that a single power-law appears to be a reasonable fit to the data. Note that the value of $\chi_{red}^{2}$ may not reflect the goodness of the fit when dealing with a small number of data points. 

\begin{table}
  \begin{tabular}{|l|c|c|c|c|c|}
  \hline
  Name& $\alpha^{con}_{1}$ & $\alpha^{var}_{1}$ &$\chi_{red}^{2}$ \\
   \hline
   \hline
   IC 694                      &2.29$\pm$0.09&2.36$\pm$0.10&0.85\\
   NGC 3690                &1.82$\pm$0.09&1.82$\pm$0.10&0.43\\
   IRAS F17578-0400   &1.74$\pm$0.11&1.78$\pm$0.07&0.82\\
   IRAS F17138-1017   & 1.53$\pm$0.12&1.64$\pm$0.08&0.58\\
   IRAS 18293-3413     &1.88$\pm$0.15 &1.89$\pm$0.09&1.49\\
   IRAS F16516-0948   &2.17$\pm$0.15 &2.41$\pm$0.07&2.17\\
   IC 883                    &1.45$\pm$0.13&1.58$\pm$0.07&0.47\\
   IRAS 19115-2124     &2.08$\pm$0.10&1.97$\pm$0.08&0.61\\
   \hline
   Average:      & 1.87$\pm$0.30 &  1.93$\pm$0.31 \\  
   \hline
\end{tabular}
\caption{\small Power-law indices from weighted linear fitting of the LFs in Fig.\,\ref{LF_fig} using the relation in Eq.\,\ref{conv_slope}. $\alpha^{con}_1$ and $\alpha^{var}_1$ are, respectively, the indices derived from binning with a constant and a variable bin width. $\chi_{red}^{2}$ show the reduced Chi Square values for the single power-law fits using the constant binning. The uncertainties  in the slopes $\alpha$ are derived from the rules of propagation of errors in Eq.\,\ref{conv_slope}, after calculating the uncertainty in $\beta$ which is the weighted linear slope shown in Eq.\,\ref{log_log_slope}. Note that in the case of a small dataset, e.g. IRAS\,F16516-0948 and  IC\,883, the value of $\chi_{red}^{2}$ may not be a good representation of the goodness of the fit. } 
\label{slope}
\end{table}

\subsection{Effect of binning on the LFs}

The shape of the LF may be affected by sample binning: for example, \citet{2005ApJ...629..873M} argued that the difference in the value of $\alpha$ can be as large as 0.3 for small datasets.  To check for consistency of our results, we constructed the LFs using two different methods: first using a constant bin size as above, and secondly using a variable bin size and assigning an {\em equal number} of objects to each bin, as proposed by \citet{2005ApJ...629..873M}.
Table\,\ref{slope} lists the values from both methods:  $\alpha^{con}_{1}$  is the slope of the LF for a constant bin size, while $\alpha^{var}_{1}$  is the one that results from using a variable bin size.  Though the values  derived from the two binning methods are slightly different, they are consistent within the error estimates of each fit.

In addition, we ran systematic tests on the effect of using a particular size of the constant-size bin. The scatter of the fitted index values $\alpha$ with a wide range of bin sizes is of the same order, $\sim 0.15$, as the uncertainties of the slope fits presented in Table\,\ref{slope}, and we conclude that bin size does not influence our final results. The characteristics of the resulting LF slopes will be discussed in $\S$\,\ref{discussion}.

\subsection{IRAS\,18293-3413: Schechter and broken power-law fits of the CLF}\label{double slope}

Figure\,\ref{LF_fig} suggests that in the  case of IRAS\,18293-3413 a two-component power-law would be a better fit to the CLF.  Hence we also fitted this LF of the target with a broken power-law with both slopes as free parameters, but the break, or bend point, $M_{bend}$ fixed at a location chosen by eye. The form of the fit is expressed as follows in a mag-logN plot: 

\begin{equation}
\small
logN(M_K)\sim \left \{ 
\begin{array}{lcl}
 \beta_3 M_K & \, {\rm for} & M_K \geq M_{\rm bend} \\
 \beta_2 M_K & \, {\rm for}&M_K < M_{\rm bend} \\
\end{array}
\right .
\label{double_equation}
\end{equation}
where  $\beta_2$ and $\beta_3$ are related, respectively, to the bright and faint slopes $\alpha_2 = 2.17 \pm 0.15$ and $\alpha_3 = 1.53 \pm 0.47$ of the CLF double power-law fits using Eq.\,\ref{conv_slope},  also plotted and labelled in Fig.\,\ref{LF_fig}  as $\alpha_2$ and $\alpha_3$. 

We also performed a Schechter fit to the CLF of the target with the form of: 
\begin{equation}
\phi (M)dM = con \times X^{\alpha_S + 1} e^{-X}dM
\end{equation}
where
\begin{equation}
X = 10^{0.4(M_K^{\star} - M)}
\end{equation}
which resulted in a characteristic magnitude of $M_K^{\star} \approx -17.4$\,mag  and a faint end slope $\alpha_S = 1.56 \pm 0.19$. Note that the values of $\alpha_3$ and $\alpha_S$ are similar for the two types of fit. 

A bent LF for IRAS\,18293-3413 fits better than a single power-law slope at the $\sim 1.5 \sigma$ level.  Though the difference is not very large, it is interesting, especially given that this galaxy has the largest statistics of SSCs in our present sample. We will discuss the case further in Section\,\ref{bentlf}.

\section{Discussion}
\label{discussion}

\subsection{The effect of blending on the LFs }
\label{blending_sec}

\begin{figure*}
\centering
\resizebox{0.33\hsize}{!}{\rotatebox{0}{\includegraphics{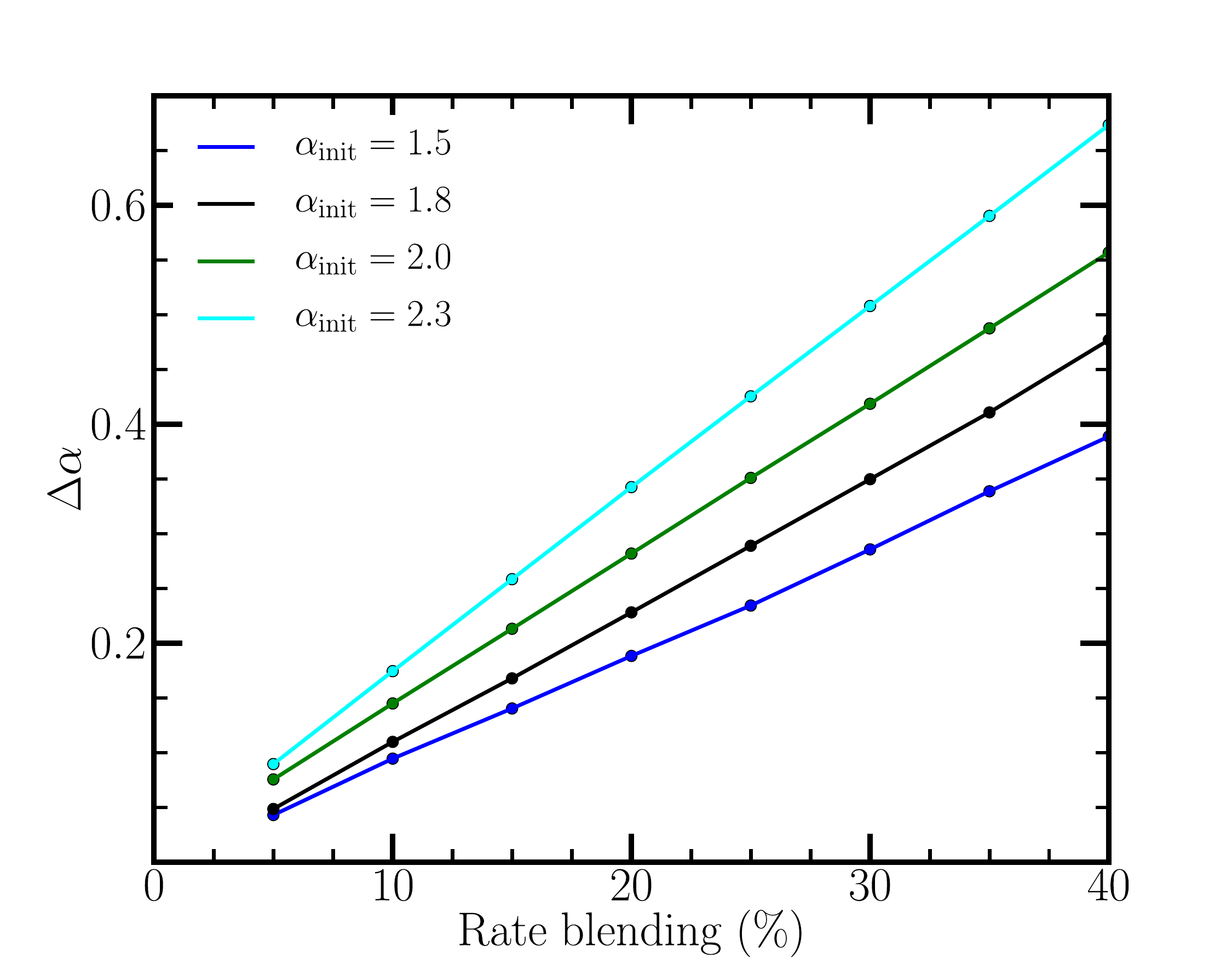}}}
\resizebox{0.33\hsize}{!}{\rotatebox{0}{\includegraphics{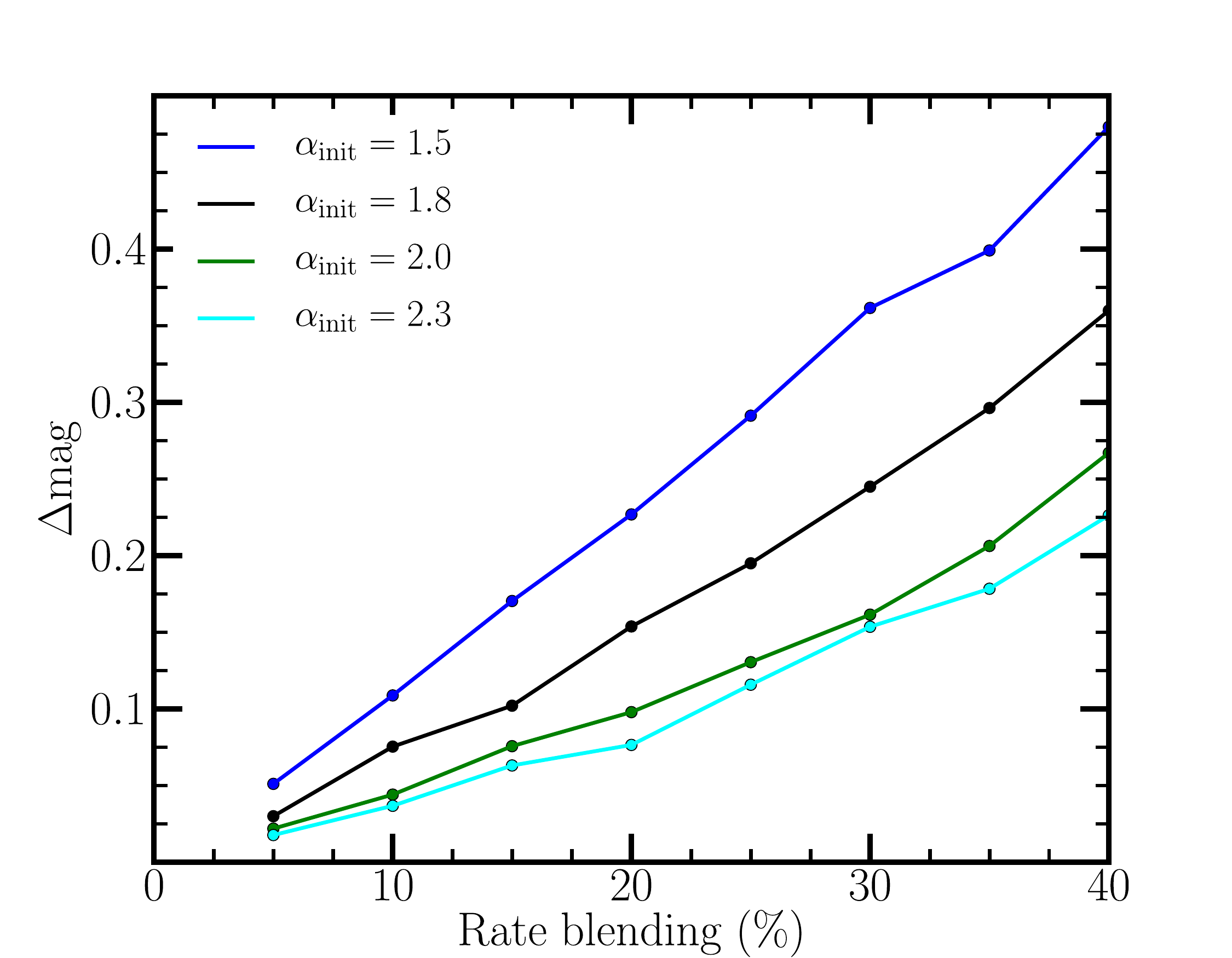}}}
\resizebox{0.33\hsize}{!}{\rotatebox{0}{\includegraphics{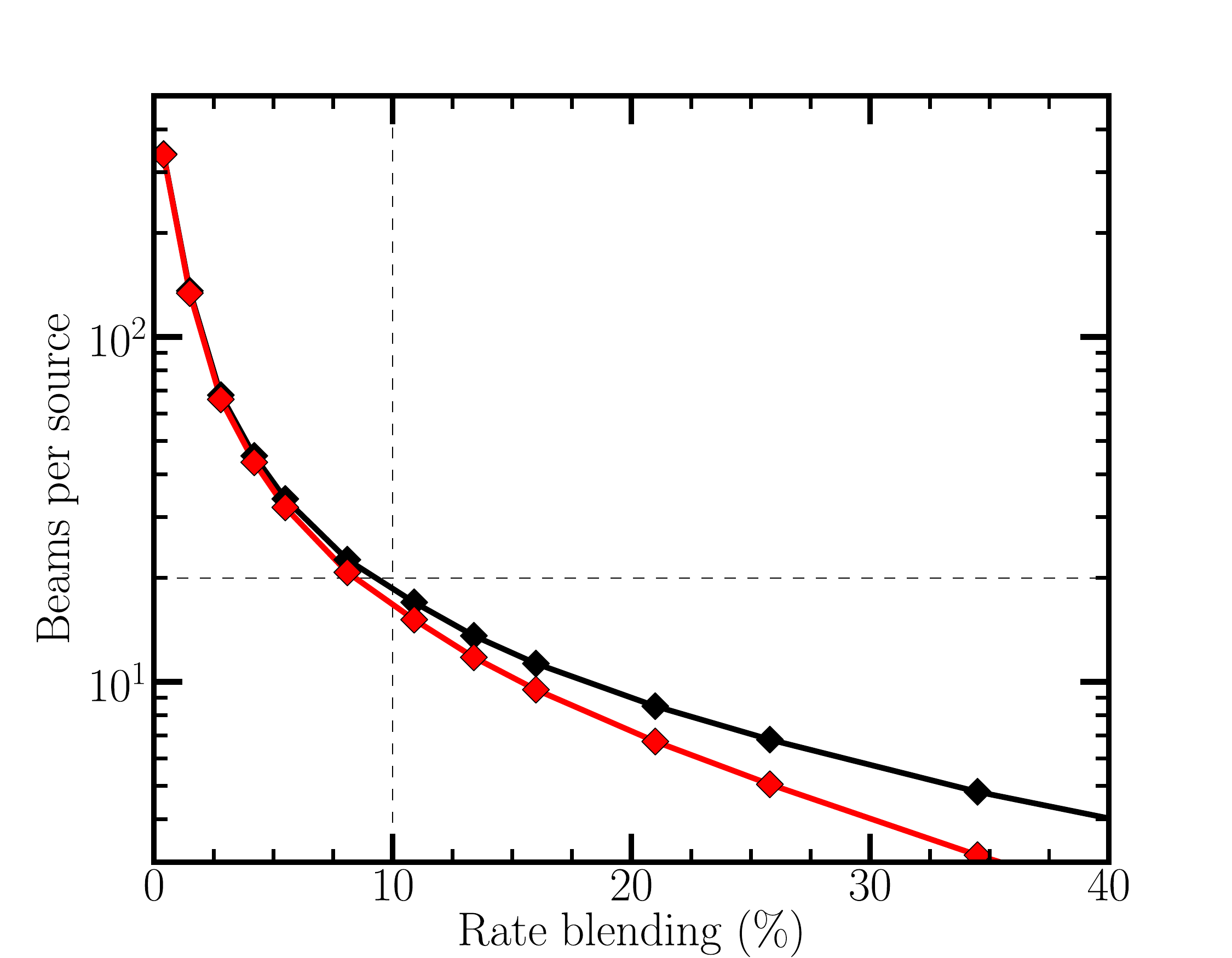}}}
\caption{\small Results from our MC blending simulation.  {\em Left}: the difference in the indices $\Delta \alpha\,=\,\alpha_{init}\,-\,\alpha_{new}$ plotted against the blending rate considering different values of $\alpha_{init}$. The figure indicates for instance that a blending rate of 8\,\% would lead to a deviation of $\sim$\,0.11 in the initial value of the index  $\alpha_{init}$\,=\,2.  {\em Middle:} The same simulations, but plotting the magnitude difference of the {\em brightest} source.   Note that intrinsic clustering characteristics are not taken into account in this simple simulation.   {\em Right:} Surface density expressed as \textgravedbl beams per source\textacutedbl\,vs.\ blending rate; see Section\,\ref{confusion} for details.  A figure of 20 \textgravedbl beams per source\textacutedbl\,is traditional confusion limit rule-of-thumb, and it is seen that at this surface density the blending rate is $\sim$\,10\,\%, resulting in $\Delta \alpha \sim 0.1$ to 0.2 depending on the initial LF slope, according to the left panel.}
\label{blending}
\end{figure*}

The distances of our sample galaxies, 45\,$<$\,$D_{L}$(Mpc)\,$<$\,100 and one at 200\,Mpc, means that any individual SSC cannot be resolved.  Our spatial resolution of $\sim$\,0.1\textacutedbl\,corresponds to a physical size of $\sim$\,20 to 40\,pc depending on the distance, and nearly 100\,pc in the case of IRAS\,19115-2124, and our photometric apertures are of comparable size.  Given that the effective radii of SSCs from the literature are in the range of 3 to\,5 pc  \citep[e.g.][]{1999AJ....118.1551W}, such aperture sizes may contain more than a single SSC candidate.  Hence, blending of SSCs and complexes of SSCs will most probably contaminate our SSC counts  despite the use of AO imaging. Is it then reasonable to even refer to these as SSCs, or should we rather talk, for example, about  \textgravedbl knots of star-formation\textacutedbl\,(\citealp{2011AJ....142...79M})?   To address this we next attempt to estimate how much blending and crowding affect our analysis.  We first perform a simulation estimating the effect of blending on the LF slopes, then examine what happens to photometry of SSCs of a nearby system when it is moved to a larger distance. We also investigate the relation between the SSC surface density and confusion limits.  In the end we conclude that blending effects are not significant within the SSC luminosity range considered, and in our case the term \textgravedbl SSC candidate\textacutedbl\,is a perfectly reasonable one for sources in our photometric catalogues.

\subsubsection{MC simulation of blending in LFs}
\label{blendsim}

We performed a MC blending simulation to quantify the effects of crowding on the values of the power-law index $\alpha$.  A random population of $N$ artificial sources was created within the same magnitude range  as our observational data ($-\,20\lesssim$\,$M_{K}$\,$\lesssim$\,$-12$) drawn from a LF with an initial index $\alpha_{init}$\,=\,2.  We randomly selected two artificial sources from this population then blended them together. This process was repeated {\it x} times until the original population had {\it x} blended sources in total, corresponding to a blending rate $y = x/N$\,\%. The blending was done step-by-step, so that a \textgravedbl new\textacutedbl\,blended source entered back into the catalogue as a new source and may be randomly blended with a third source, etc.  A new power-law index, $\alpha_{new}$ was determined from a fit to the blended source luminosity distribution.  The difference in the slopes, $\Delta \alpha\,=\,\alpha_{init}\,-\,\alpha_{new}$ was then determined.  We ran the simulation with {\it y} ranging from 5 to 60 per cent in steps of 5\,\%.  A particular blending realisation was repeated 1000 times for each blending rate.   We also ran the MC simulation for $\alpha_{init}$\,=\,\{1.5, 1.8, 2.3\}.  We found that blends of 3 or more sources are rare below $\sim$\,20\,\% blending rates.  

Figure\,\ref{blending}, left panel, plots $\Delta \alpha$ against the blending rate with different values of $\alpha_{init}$.  The sense is always that of {\it flattening} of the slope.  A  blending rate of 10\,\%  for example would lead to a deviation of $\Delta \alpha \sim$\,0.14 from $\alpha_{init}$\,=\,2.  The figure indicates that to get significant flattening of $\alpha$ by 0.3 or more, one would need blending rates of $\sim 25$\,\% or higher for an initial slope of 2, or nearly 40\,\% for the slope to flatten from 2 to 1.5.  The blending rates needed are higher for a steeper initial slope and lower for a flatter initial slope as expected.  The blending rate is obviously related to the surface density of objects in the field, and we will return to the likely real blending rates of our targets in Section~\ref{blendeffect}.

Figure\,\ref{blending}, middle panel, plots the change in magnitude of the brightest object; with more blending there is a greater chance that the brightest object is in fact a blend. Again we see that significant changes, $\Delta {\rm mag}  > 0.5$, in the magnitude require blending rates of 40\,\% or higher. 

The simulation does not take into account varying degrees of blending, nor changes in completeness limit.  Most significantly, however, the simulation implicitly assumes a random surface distribution of targets, whereas SSCs are clearly clustered in (most) galaxies.  While the simulation gives a good feel for the expected levels of changes in LF slopes due to blending, a more realistic estimate should be based on real data.

\subsubsection{SSCs in a redshifted Antennae}

\begin{figure*}
\centering
\resizebox{0.48\hsize}{!}{\rotatebox{0}{\includegraphics{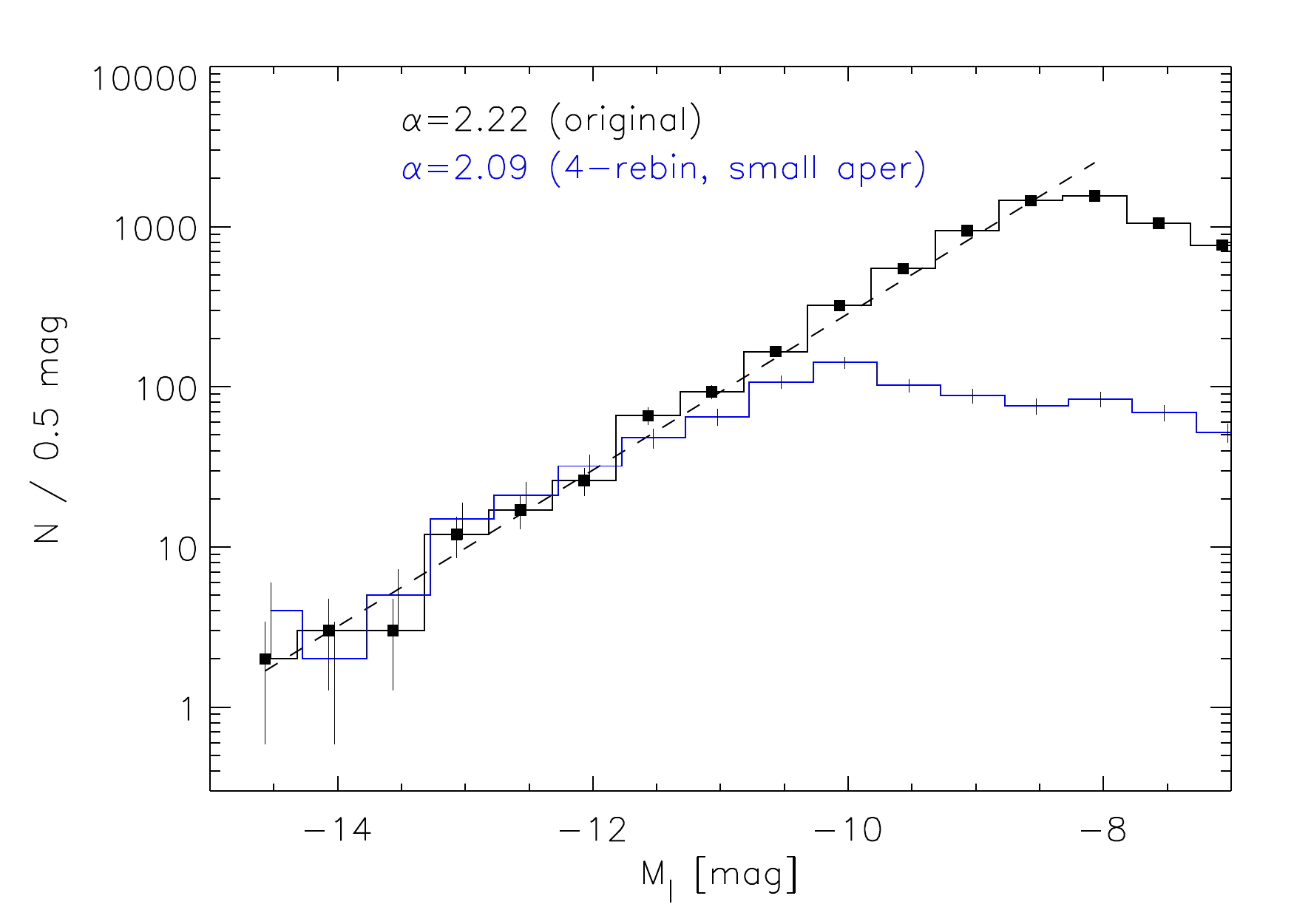}}}
\resizebox{0.48\hsize}{!}{\rotatebox{0}{\includegraphics{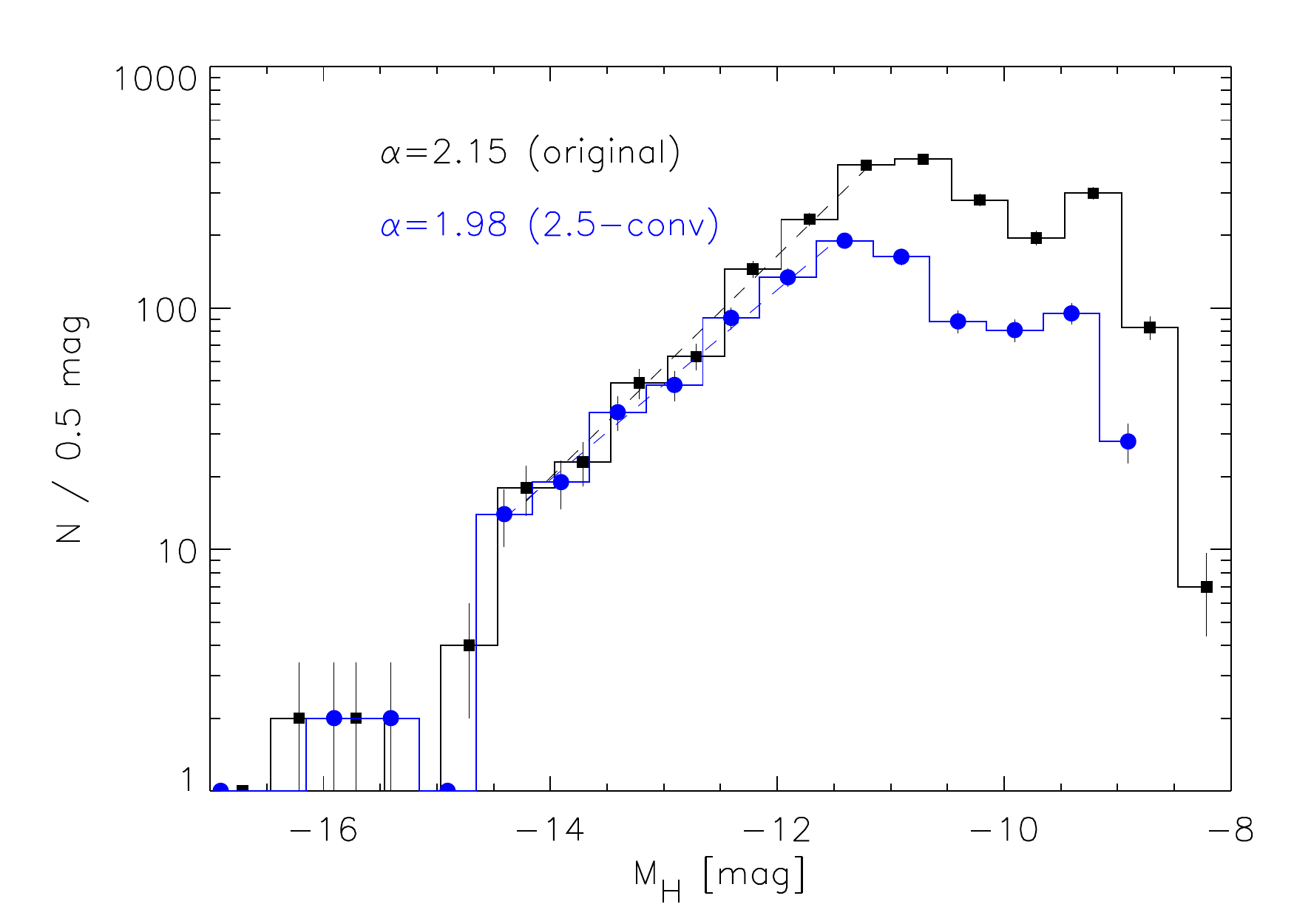}}}
\caption{\small LFs of SSCs derived from $HST$ images of the Antennae galaxies.  {\em Left:}  The black histogram shows the LF from the original $I$-band (F814W) image, and the blue after convolving and rebinning the image to correspond to a 4 times larger distance ($\approx84$\,Mpc).  The best-fit power-law slopes are indicated.  {\em Right:} The same test for the $H$-band (F160W) image, but now the comparison is made so that the convolved image matches both the PSF size and the pixel size of our IRAS\,18293-3413 NACO image.  In both tests only mild flattening of the LF slope is observed.
}
\label{ant_lf}
\end{figure*}

The Antennae system, at $D_{L}$\,$\sim$\,20\,Mpc, is a popular laboratory for SSC studies (\citealp[e.g.][]{1999AJ....118.1551W, 2001ApJ...561..727Z}),  and we use its well-studied star cluster population for a blending estimate.  In particular, we retrieved $HST$/WFC3 images (PI: Whitmore) in the UVIS/F814W ($I$-band) and IR/F160W ($H$-band) filters from the Hubble Legacy Archives.  Adopting a distance of 22\,Mpc the 0.04\textacutedbl\,pixels in the $I$-band translate to a 4.3\,pc physical size and the PSF to $\sim$\,8\,pc.  In the $H$-band dataset the pixel size is equivalent to 9.6\,pc and the PSF to $\sim$\,16\,pc.   

We first convolved and rebinned the $I$-band Antennae image to correspond to being 4 times further away.   At 88\,Mpc it is well within the distance range of our sample (Table\,\ref{coords_table}), and the PSF resolution element of 32\,pc is also similar to that in our NACO and ALTAIR/NIRI images.  SSCs were detected both from the original and convolved images. We do not take into account the change in detection limits, so as to test only the effects of blending.   The photometry and LF constructions were done similarly to our AO data described above, except that we do not attempt completeness corrections here as they are irrelevant to the main goal.  Following other $HST$ studies of the Antennae \citep[e.g.][]{2010AJ....140...75W} an aperture radius of 0.1\textacutedbl\,was used on the original (unbinned) image, corresponding to 10.8\,pc. Nominal photometric zero-points and aperture corrections were taken from the WFC3 on-line manuals. The resulting LF is shown as the black points in Fig.\,\ref{ant_lf},  left panel, and a single power-law slope of $\alpha = 2.22$ was found; the fit is performed down to  $M_I \approx -8.5$\,mag where incompleteness appears to be setting in.   This value is in fact very consistent with other studies though no completeness or stellar contamination corrections were made in our test, for example \citet{2010AJ....140...75W}  found $\alpha = 2.26$ in the $V$-band.

The source extractions and photometry were done in the same way for the \textgravedbl redshifted\textacutedbl convolved image.  An additional aperture correction is needed however since the PSF and pixel characteristics change, and the new correction was simply determined by matching the extracted magnitudes of a handful of foreground stars in the different images and apertures.  The LF shown in blue in Fig.\,\ref{ant_lf} results from extractions from the convolved image when using a small 0.75 pixel aperture, corresponding to 13\,pc radius. As expected, the completeness limit is some 2 magnitudes  brighter due to blending only, while the LF slope becomes only slightly flatter at $\alpha=2.09$.  The aperture size used does not have an effect on the slope, but we will return to this aspect in more detail below.

We also examined what happens to the 50 and 200 brightest original SSCs in the convolved image.  Of the 50 brightest ones 7 were not detected as individual objects after the \textgravedbl redshifting\textacutedbl (14\,\% blending rate)  while 53 of the 200 brightest ones were not detected (27\,\% blending).  In the MC simulation of Sec.~\ref{blendsim} these blending rates would have resulted in a flattening of the slope by $\Delta \alpha \approx 0.20$ and $\approx$\,0.37, respectively.  These numbers are somewhat larger than the measured flattening in the range 0.1 to 0.15 of the convolved image LF slope in the test above (Fig.\,\ref{ant_lf}).  The recovered magnitudes of the brightest 50 SSCs are a mere $\sim0.1$\,mag brighter than in the original image.

In case clustering properties of SSCs would be different in the optical compared to NIR, we repeated a similar test using the $H$-band image of the Antennae.  This time we modified the original $HST$ image to match both the physical pixel scale and PSF size of the NACO IRAS\,18293-3413 image at $D_L \approx 75$\,Mpc.  Rebinning was not necessary because of the smaller pixels of the NACO instrument, while a factor of 2.5 widening of the PSF was performed.  Magnitudes were measured in 20\,pc radii in the original image, matching the aperture used in the  \citet[e.g.][]{2010AJ....140...75W}  NICMOS data, and 30\,pc radii in the redshifted case, as done for the NACO IRAS\,18293-3413 images.  The resulting LFs are shown in Fig.\,\ref{ant_lf}, right panel, indicating a flattening of the slope by 0.15, very similar to what was found for the $I$-band $HST$ image.  The magnitudes do not change significantly this time either,  the average difference being less than 0.1\,mag for the 200 brightest SSCs. 

To understand and differentiate between the effects of spatial resolution and photometric aperture used with a given resolution, we redid the tests above  with different convolutions using PSFs ranging from 10 to 100\,pc, as well as using numerous aperture sizes in the same range.  Some results become clear.  First of all, the photometric aperture used at a given resolution  {\em does not change the slope significantly}.   Variations of $\sim 0.1$ were typically found for the value of $\alpha$.  
Secondly, the LF does however shift to brighter magnitudes as the apertures grow.  To recover as closely as possible the intrinsic SSC counts {\em the smallest possible aperture should be used}, assuming a reliable aperture correction can still be determined. With the largest tested apertures the bright SSCs brighten by nearly a magnitude, while using apertures smaller than about $20-30$\,pc radius, the brightening stays within typical photometric errors of $\sim 0.2$\,mag. 
Thirdly, when analysing the effects of different convolutions we found that the resulting LF slope remains relatively unchanged, staying within $\Delta \alpha = 0.2$ of the original slopes of 2.15 and 2.22 ($H$ and $I$-band respectively) until sizes of about 40\,pc are reached, after which the slopes flatten rapidly  reaching $\alpha \sim 1.5$ at the poorest resolutions of a 100\,pc physical size.   

\begin{figure*}
\centering
\resizebox{0.47\hsize}{!}{\rotatebox{0}{\includegraphics{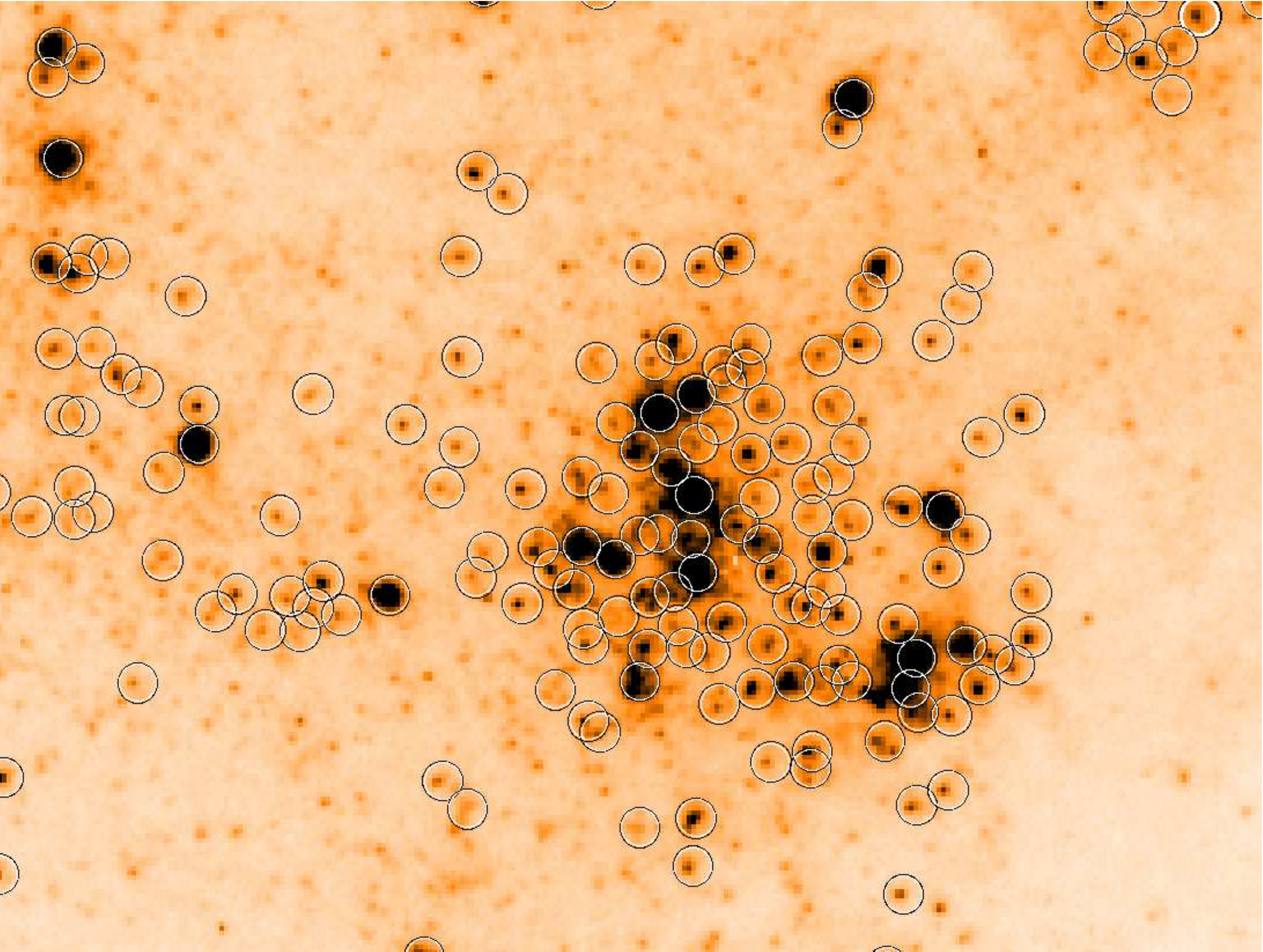}}}
\resizebox{0.47\hsize}{!}{\rotatebox{0}{\includegraphics{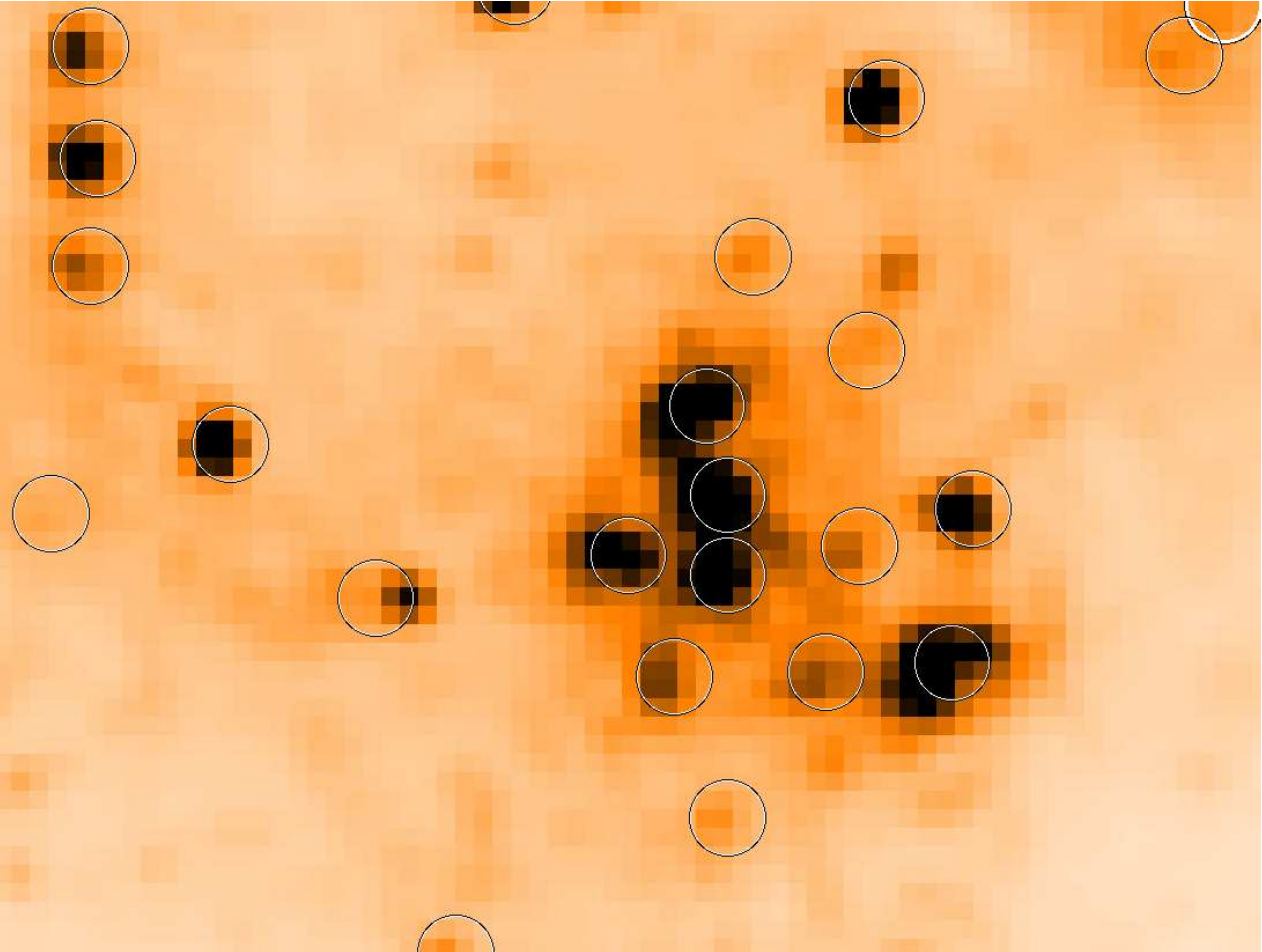}}}
\caption{\small A 10\textacutedbl\,by 8\textacutedbl (1.1 by  0.9\,kpc) region in the $HST$ $I$-band image within the Antennae (region E in \citealp{2010AJ....140...75W}). {\em Left:}  The original image with $\sim 160$ SSC candidates detected.  {\em Right:} The image after taking the galaxy four times further away. Only $\sim 20$ SSCs are now detected due to blending, though the brighter population remains relatively unchanged.   }
\label{example_dets}
\end{figure*}

Figure\,\ref{example_dets} shows an example of a small region within the Antennae in both the original and convolved images of the $HST$ $I$-band which illustrates these effects. While numerous faint SSCs disappear or are blended into bright SSCs, the latter are generally recovered in the convolved image with close to their proper magnitudes, unless located in very crowded regions.   For example the more isolated bright SSCs on the edges of the image are recovered within $\sim 0.1$\,mag of the original magnitudes, while the three bright SSCs in the middle of the central conglomeration have brightened by about 0.4\,mag. The largest changes to the photometry are in cases where SSCs of similar brightness happen to lie near each other and blend into a single significantly brighter object, such as the bright SSC at bottom right which has blended into one source 1.1\,mag brighter than either of the originals.  How severe these effects are naturally depends upon on the overall surface density of objects, which is discussed next. 

\subsubsection{The effects of confusion limits}
\label{confusion}

Historically, blending and crowding as discussed above are referred to as \textgravedbl confusion\textacutedbl\,especially in the radio and far-IR studies, and in particular \textgravedbl bright source confusion\textacutedbl. When a confusion limit is reached depends, in addition to the shape and size of the spatial resolution element, on the slope of the source counts.  With very steep slopes {\em confusion noise} dominates, i.e. the many undetected sources at and below a detection limit give rise to an effective noise-level (see e.g. \citealp{2001MNRAS.325.1241V} and references therein for full discussions) that deeper observations cannot penetrate.  With much flatter slopes the bright sources lying too close together for satisfactory extraction at a given spatial resolution tend to dominate.  The SSC source count slopes are closer to this latter regime, but it is prudent to search for this surface density-related confusion limit also in cases where the images are not truly confusion-limited yet.   

A 20 to 40 \textgravedbl beams per source\textacutedbl\,confusion limit is often used as a rule-of-thumb. For example, for a resolution element FWHM\,=\,0.1\textacutedbl,  setting the beam size as 
\[ \Omega = \pi \, \frac{FWHM^2}{4 \, \ln(2) },  \]
then 20 beams per source corresponds to having one SSC per every 0.47\textacutedbl $\times$ 0.47\textacutedbl\,region.  Hence, having over 500 sources within the area of IRAS\,18293-3413 (see Fig.\,\ref{SSC_18})  would mean reaching the limit, while we detected $\sim$\,200. However, in the core regions the surface density is definitely approaching the confusion limit.  

\begin{table}
  \centering
  \begin{tabular}{lccccccc}
  \hline
   \hline
 & \multicolumn{6}{c}{Distance [Mpc]}   \\
   &  10 & 20  & 40 & 80  & 120 & 200 \\

\multicolumn{2}{l}{PSF [\textacutedbl]}   &   &   &   &  &  \\

 \hline
     0.05\textacutedbl  & 440 (1900) &  440 (480)  & 220  &  130  &  58 &   22 \\
    0.075\textacutedbl  & 440 (850)   & 220   & 98  &  59    & 26  &   10 \\ 
     0.10\textacutedbl  & 440 (480)   & 120   & 55  &  33    & 15  &   5.6 \\
     0.20\textacutedbl   & 120    & 30   & 14   &  8.3    & 3.7   & 1.4 \\
     0.30\textacutedbl    & 53   & 13    &  6.1   &  3.7   &  1.6     &0.62 \\
     0.50\textacutedbl    & 19    & 4.8  &   2.2   &  1.3  &   0.59    & 0.22 \\
     1.00\textacutedbl    & 4.8    & 1.2 &    0.55   &  0.33  &   0.15   &  0.06\\
   \hline
\end{tabular}
  \caption{\small Surface densities, in units of kpc$^{-2}$, which would result in {\em confusion-limited} observations of SSCs, when the limit is defined as 20\,\textgravedbl beams per source\textacutedbl.   For small distances/PSFs the assumed 10\,pc physical size of SSCs is {\em resolved} and the value corresponds to this confusion limit;  the value corresponding to the actual resolution element is given in parentheses.
 }
\label{confvalues}
\end{table}

It is worthwhile to define a general confusion limit in terms of physical sizes for SSCs. Taking SSCs to be of size 10\,pc, the 20 beams-per-source criterion corresponds to a SSC density of 440\,kpc$^{-2}$.   If SSCs are seen more densely packed than this then better resolution is not likely to help extend SSC detection.  
Over the approximately $9 \times 9$\,kpc$^2$ size of the Antennae, for example, this would mean some 40000 sources; extrapolating the counts we extracted from the $HST$ image, this level would be reached around $M_I \sim -6.5$\,mag.  As seen above, in individual {\em regions} of the galaxy system, the confusion limit must be reached at much brighter magnitudes since the SSCs are clustered. This is reflected in the fact that significant incompleteness starts appearing already at $M_I \sim -8$\,mag in our test (Fig.\,\ref{ant_lf}, left panel). 
{\em If} the $HST$ image was truly confusion-limited, and assuming a LF with $\alpha=2$, the completeness limit would be expected to brighten by 3 magnitudes after the factor 4 convolution, whereas a change of 2\,mag was observed.   

We list in Table\,\ref{confvalues}  {\em physical} surface densities per kpc$^2$ at which the confusion limit is likely to be reached at a given distance and spatial resolution, given the definitions above.   With the typical PSF of $\sim 0.1$\textacutedbl, the distances of our sample, and the number of detected SSCs, only the core regions of IRAS\,18293-3413 come anywhere close to these confusion limits, as well as  IRAS\,19115-2124 due to its distance.  This is not to say that individual SSCs will not blend of course, especially in clustered star-forming regions of the galaxies as shown in the tests based on the Antennae. For example, the surface density of detected objects in the left panel of Figure\,\ref{example_dets} is $\sim 160$\,kpc$^{-2}$.  At a distance of $\sim 20$\,Mpc and for a resolution element of 0.075\textacutedbl this is not quite yet confusion-limited according to Table\,\ref{confvalues}, though it is approaching it.  This is also seen from the corresponding $78$ beams-per-source surface density.  However, within the clustered sub-region the equivalent values would be $\sim 500$\,kpc$^{-2}$ and $\sim 25$ beams-per-source, respectively, i.e. that region is confusion-limited.  

To see how the surface densities quantitatively relate to the LF slope changes, we ran another MC simulation adding increasing amounts of randomly distributed equally bright sources in an otherwise empty frame and extracting them with {\tt SExtractor} using typical parameters.  This was done hundreds of times at several surface densities, and the fraction of unrecovered sources is simply the {\em blending rate} at each surface density. The change in LF slope with a given intrinsic LF shape was already simulated in Section\,\ref{blendsim} as a function of this blending rate.  The right panel of Fig.\,\ref{blending} connects the two by plotting the \textgravedbl beams per source\textacutedbl\,surface density against the blending rate.  The black curve is the intrinsic surface density of the simulation, and the red curve is calculated from the extracted surface density.  An {\em observed} surface density of objects can be used to get an expected blending rate using the red curve (for completeness-corrected counts the black curve is more appropriate) and this can then be converted to a likely $\Delta \alpha$ value of the LF.  While the vertical displacement of the surface density vs.  blending rate curves will depend on the source detection algorithm and clustering of objects, our tests show that the relations do serve as a realistic approximation of the quantitative effects involved.

\begin{figure*}
\centering
\resizebox{0.47\hsize}{!}{\rotatebox{0}{\includegraphics{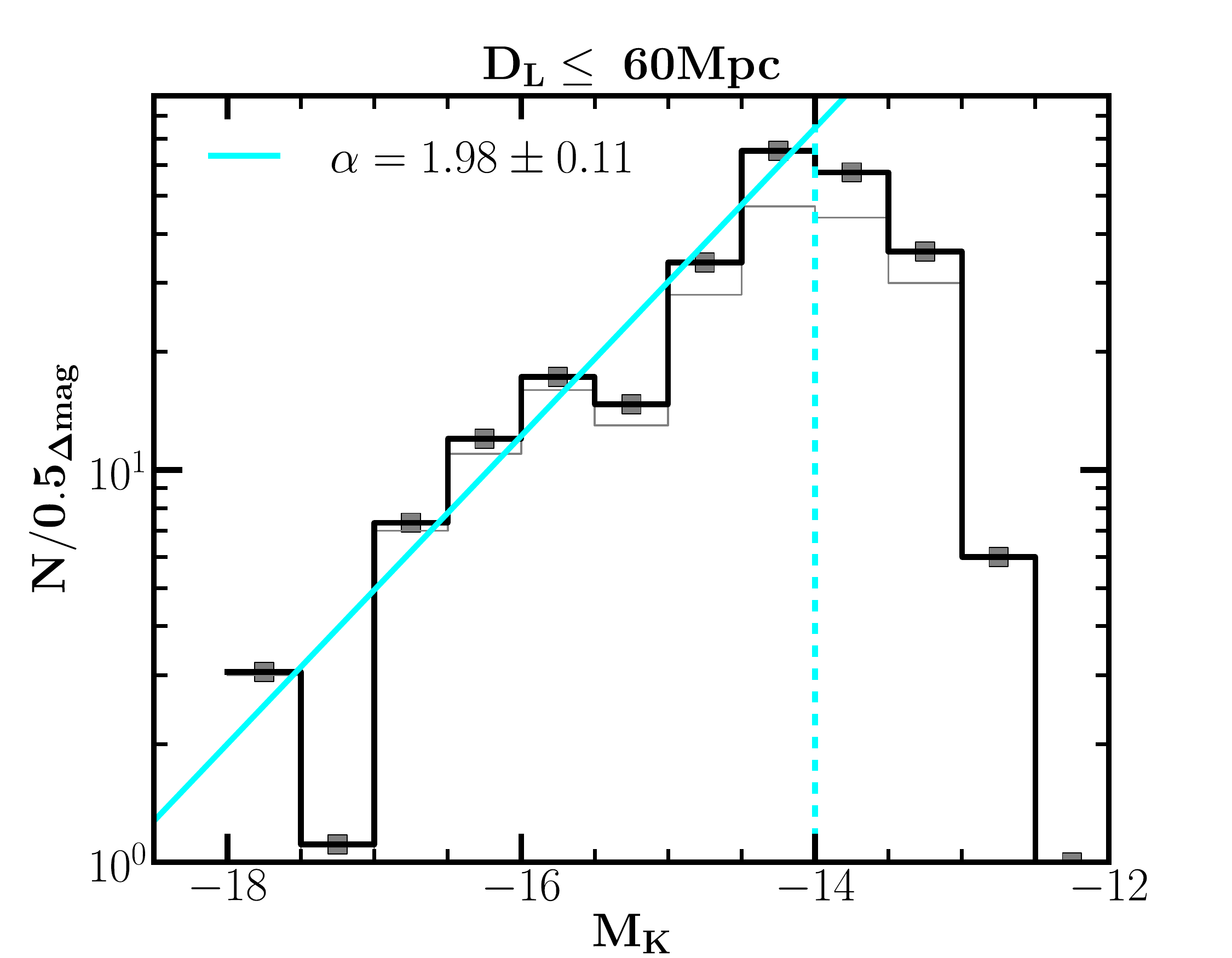}}}
\resizebox{0.47\hsize}{!}{\rotatebox{0}{\includegraphics{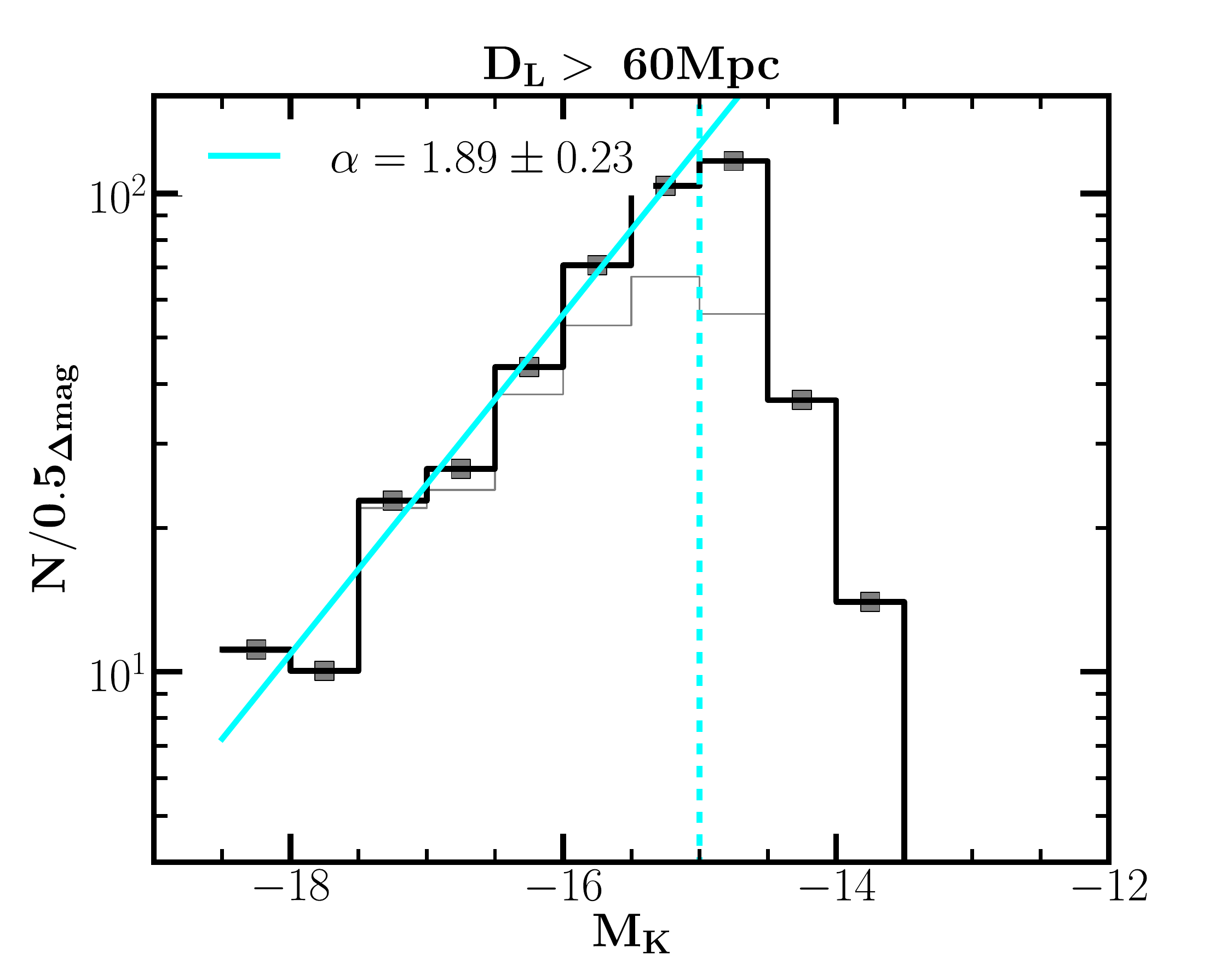}}}
\caption{\small SSC LFs of two subsamples generated from our observational data but segregated by distance. The dataset from IRAS\,19115-2124 is not included to avoid bias in the analysis. {\em Left:} SSC LF of the closer subsample where the targets have distances $\leq$\,60\,Mpc. {\em Right:} SSC LF of the more distant targets, excluding IRAS\,19115-2124. The values of the slopes appear to be consistent within the uncertainties.}
\label{LF_distance}
\end{figure*}

\subsubsection{Effect of blending in our sample}
\label{blendeffect}

The average surface density of detected SSC candidates in our target galaxies ranges from a low of 0.3\,kpc$^{-2}$  in IC\,694, or 168 beams per source, to a high of 0.8\,kpc$^{-2}$ in the case of IRAS\,18293-3413, or 87 beams-per-source.  According to Fig.\,\ref{blending}, the corresponding blending rates are well below $5$\,\%, meaning that $\Delta \alpha < 0.1$ for any LF shape, most likely $<0.05$ for most of the targets.  Because of the uncertainties involved, such as the difficulties of properly accounting for clustering, we prefer not to attach $\Delta \alpha$ values for each individual target, but rather stress that these {\em must} be less than 0.1.  However, in our most extreme case of the {\em central} $\sim 2\times2$\,kpc region of IRAS\,18293-3413, the SSC surface density equates to 20--30 beams per source, and $\Delta \alpha \sim 0.15$ would be expected in that region.

As a further check, Fig.\,\ref{LF_distance} shows separately the LFs for SSCs out to distances $D_{L} \leq 60$\,Mpc, and for $D_{L} > 60$\,Mpc (excluding IRAS\,19115-2124 since it is so much further away).  If blending was a serious issue in our sample, the more distant subsample would be expected to show a flatter slope. The single power-law indices of the LFs below and above 60\,Mpc are $\alpha = 1.98 \pm 0.11$ and  $\alpha = 1.89 \pm 0.23$ respectively, i.e. there is no significant difference in the LF slope, nor in the numbers of detected SSCs, with distance.

In summary, we are confident that in the luminosity range of interest, $M_K < -14$\,mag, any flattening of LF slopes due to blending is below $\Delta \alpha \sim 0.1$.  With the possible exception of IRAS\,19115-2124 where the spatial resolution is just a physical size of 90\,pc, the luminosities of detected $K$-band point sources will be dominated by a single bright SSC rather than whole knots of star formation.  In the case of Antennae-like SSC populations, blending and crowding do flatten the LF slope, but significantly so only at resolutions poorer than a 40\,pc physical size. The photometric apertures used should be  as small as possible to recover intrinsic luminosities, though the aperture does not have a significant effect on the LF shape.  Assuming a 10\,pc scale for SSCs,  the confusion limit is reached at a surface density of 440 SSCs per kpc$^2$, or less when clustering and larger resolution elements are a factor (Table\,\ref{confvalues}).

\subsection{Interpreting the values of $\alpha$}

The results from our fitting procedure in $\S$\,\ref{fit_LF} show that a single power-law distribution is a reasonable approximation for the young stellar cluster LFs in our sample.  
A single power-law SSC LF has been reported by many theoretical and observational studies 
\citep[e.g.][]{1997ApJ...480..235E,2003ARA&A..41...57L,2006A&A...450..129G}, 
and such behaviour may be linked to the shape and characteristics of the physically more interesting 
underlying (initial) mass function and/or the evolution and disruption of the SSC population.
As listed in Table\,\ref{slope}, the fitted single power-law indices range from   $\alpha \sim$\,1.5 to 2.4, with average single slope values of $\alpha^{con} = 1.87 \pm 0.30$ for a constant bin size, and  $\alpha^{var} = 1.93 \pm 0.23$ for a variable bin size.  Values of $\alpha$  reported in the literature  vary widely from 1.7 to 2.4 (e.g. \citealp{1997ApJ...480..235E}, \citealp{1999AJ....118.1551W}, \citealp{2002AJ....123.1381E}), with slopes in normal spirals often in the upper part of this range.  Our $\alpha$ values tend to be in the less steep part of the range, and in at least one case (IRAS\,18293-3413) either a broken power-law or a Schechter function would yield a better fit.  These two issues are discussed next.

\subsubsection{LF slopes: a difference between LIRGs and normal spirals?}

\begin{figure*}
\centering
\resizebox{0.47\hsize}{!}{\rotatebox{0}{\includegraphics{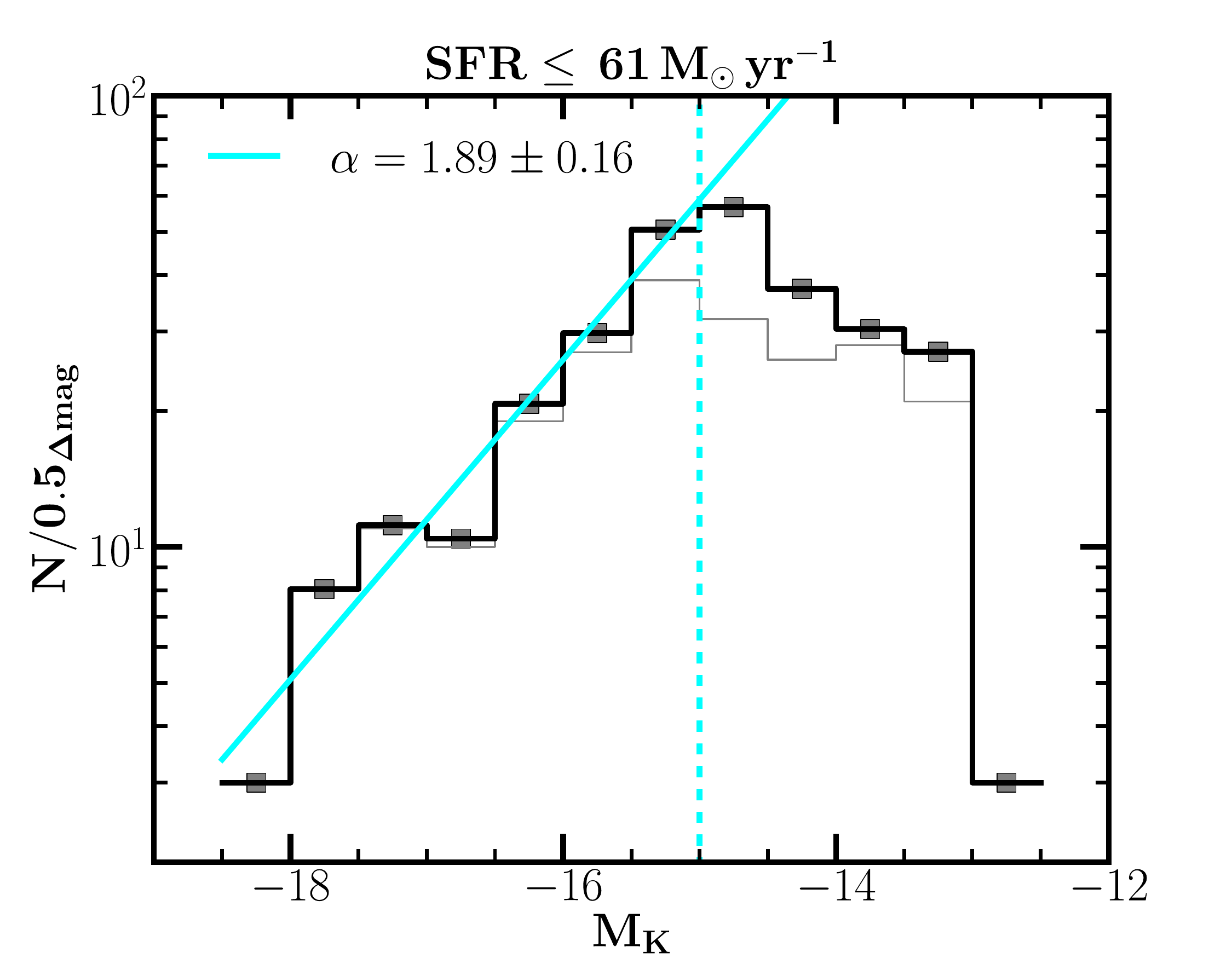}}}
\resizebox{0.47\hsize}{!}{\rotatebox{0}{\includegraphics{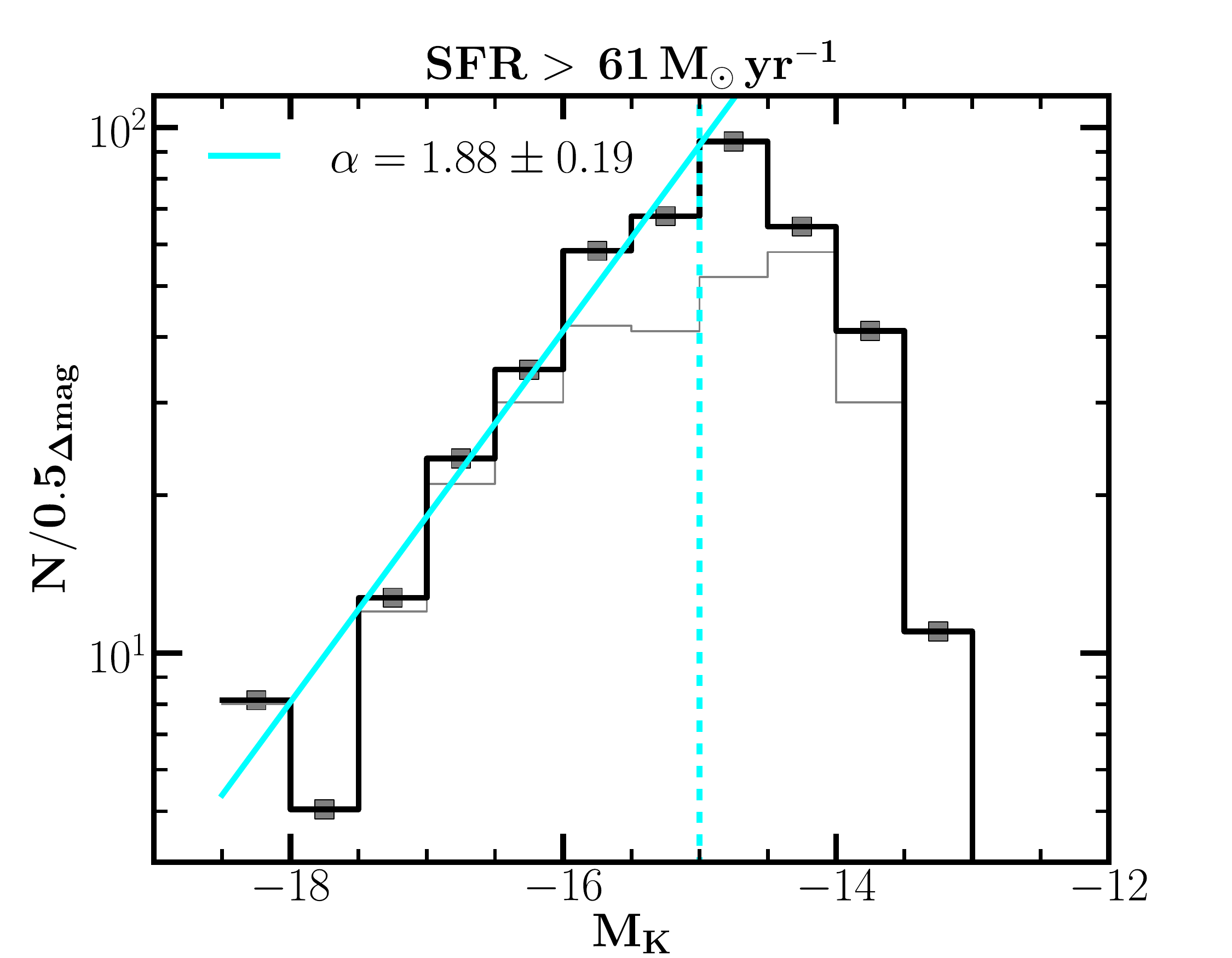}}}
\caption{\small SSC LFs of two subsamples generated from our observational data, excluding IRAS\,19115-2124, but broken by the average SFR. {\em Left:} SSC LF of the subsample where the targets have SFR $\leq$\,61\,$M_{\odot}yr^{-1}$. {\em Right:} SSC LF of the other set of targets with high SFRs. A power-law fit until the 80\,\% completeness level generates a quite similar value of $\alpha$ in both cases.}
\label{LF_SFR}
\end{figure*}

Improper incompleteness corrections can produce artificially flattened slopes.  We note however that the slopes are fitted to a range where at most 20\,\% of SSCs are missed, and where the corrections are still reliable.  The most significant change is likely to be caused by blending of SSCs but as discussed in the previous section, this is expected to be  $\Delta \alpha \approx 0.1$ at most for our targets, and likely less.  Also, we did not see any significant change in the $\alpha$ values with distance to the galaxies, which would have been expected if significant blending was present (Fig.\,\ref{LF_distance}).  
Therefore, we conclude that the intrinsic SSC LF slopes, the combined and averaged LFs having $\alpha \approx 1.82 \pm 0.25$ and $1.87 \pm 0.30$, respectively, are not affected by these observational effects, with any systematic effects being outweighed by  the statistical uncertainties.
 
 Normal spiral galaxies tend to have power-law indices in the range of $\alpha \sim 2$ to 2.4.  Calculating the average from SSC LF studies which include several spirals each \citep{2002AJ....124.1393L, 2006A&A...450..129G, 2009A&A...501..949M}, we get $\alpha \approx 2.17 \pm 0.24$ from 10 spiral galaxies.  If we include other individual galaxy studies \citep{1996AJ....112.1839S, 1999AJ....118.2071B, 1999AJ....118..777E, 2002AJ....123.1381E, 2009A&A...503...87C, 2010AJ....139.1369P, 2010AJ....140...75W}, where some targets are starbursts, though not yet LIRGs, we get an average of $\alpha \approx 2.18 \pm 0.19$ from a total of 19 galaxies. Note that the values come from different filters, mostly from $V$- and $I$-bands; if several bands were used, we chose the reddest band.  As discussed earlier the well-studied interacting Antennae system also has $\alpha \sim 2.2$ \citep{2010AJ....140...75W}.   
 
 While most of the previous SSC LF determinations in the literature, including the Antennae, have been in actively star-forming galaxies, the vast majority of them have milder SFRs than LIRGs.   Our average power-law slope for LIRGs, $\alpha \approx 1.9$, appears flatter than the slope in normal galaxies. The large sample of LIRGs studied in \citet{2011PhDT.........8V} shows an average $\alpha \sim1.8$, though blending effects there might still need to be explored.  Similarly, \citet{2011AJ....142...79M} find $\alpha \sim 1.9$ for their (U)LIRGs at $<100$~Mpc, and even flatter values at certain interaction stages of LIRGs.  Moreover, \citet[]{2010MNRAS.407..870A, 2011MNRAS.415.2388A} recently found flatter power-law slopes while probing the star cluster properties in Haro 11 and Mrk 930.   Both of the targets are blue compact dwarfs with intense star formation, and the former can also be classified as a LIRG.  Is it the case therefore that LFs are systematically flatter in extreme SF cases? 

 To see if there is a trend with SFR {\em within} our sample, Fig.\,\ref{LF_SFR} shows LFs separately for SSCs with host galaxy SFRs less than, or greater than the average $\overline{\rm SFR}$\,$\sim$\,61\,$M_{\odot}yr^{-1}$, again with IRAS\,19115-2124 excluded. The values of the slopes are $\alpha = 1.89 \pm 0.16$ for the subsample below the average and $\alpha = 1.88 \pm 0.19$ for those above.  Thus, while we do not find a correlation between the LF slope and the SFR within our present small sample, our results together with other recent studies suggest that there is a real difference between the SSC LF slopes of LIRGs and those of more quiescent galaxies.  This needs to be verified with larger samples for better statistics, and with careful blending analysis since larger samples will by necessity involve more distant targets.

Apart from effects related to the observations themselves, perhaps the most fundamental cause for flattened LFs, or breaks/bends in the LF for that matter, would come from mass- and/or age-dependent cluster disruption, as well as from differences in the cluster formation with different environments.
Young star clusters are most vulnerable to disruption, leading to variation of the integrated LF as time passes. Since LFs are the integrated sum of the distributions of individual initial LFs of SSCs of different ages and masses, any selective (e.g. mass-dependent) disruption of them would be seen as changes in the (integral) total SSC LFs \citep[e.g.][]{2005A&A...443...41M, 2006A&A...450..129G, 2007ChJAA...7..155D}.   
 \citet{2012MNRAS.421.1927K} argue that cluster {\em formation} efficiencies decrease in galaxies with higher SFRs due to increased tidal disruption in the extreme environments (\textgravedbl cruel cradle effect\textacutedbl), which might lead to flatter LF slopes if lower mass SSC formation is more affected.  Alternatively, high SFRs may favour the formation of more massive GMCs \citep[e.g.][]{2012ApJ...750..136W}.    However it is impossible to conclude whether disruption is  the main cause of the flatter slopes until we have analysed data from other filters to be able to derive LFs and MFs in different mass and age regimes.

\subsubsection{ IRAS\,18293-3413: What could a broken power-law reflect? }
\label{bentlf}

In the case  of IRAS\,18293-3413 fitting two independent slopes produces better Chi Square values ($\chi_{red}^{2} =  0.57$) than a single power-law.  The break appears to be at $M_K \approx -16.5$\,mag and persists regardless of how the data is binned.   We note that the distribution can also be well fit with a Schechter function, giving a characteristic magnitude of $M_K^{\star} \approx -17.4$\,mag ($\chi_{red}^{2} =  0.64$).
 In general, any bends or cutoffs found close to the detection or completeness limits should be viewed with caution, but in this case the break is more than a magnitude above the 80\,\%  completeness limit.

Bends have been argued to result from physical truncations of the SSC high-mass end of the SSC mass functions  or from disruptions of the less-massive SSC population (see e.g. \citealp{2006A&A...450..129G}, \citealp{2008A&A...487..937H}).  The former results in a steepening of the bright end of the LFs if the truncation is a physical one and not merely a result of a size-of-sample effect;  the latter affects the faint end of the LF by making it flatter.  Thus both can result in bends in the shape of the LF, making for better fits with broken power-laws or  Schechter functions.  The absolute magnitude of the break point would increase with the maximum mass of the SSC population, though the exact relation is a complex function of the disruption models.  The break is expected to occur at brighter magnitudes at redder wavelengths because of the expected red colour of the aged SSCs.  Whether or not these bends and truncations are in fact seen in LFs or mass functions is still a matter of intense debate (see e.g.  \citealp{2006A&A...450..129G}, \citealp{2009A&A...494..539L}, \citealp{2010AJ....140...75W},  \citealp{2012ApJ...752...96F} and references therein).

Bends in SSC LFs have typically been found in the range of $M \sim - 9$ to $- 11$\,mag in $VRI$ filters (e.g. \citealp{1999AJ....118.1551W, 2006A&A...450..129G, 2006A&A...446L...9G, 2010MNRAS.405.1293S, 
2010MNRAS.407..870A, 2011MNRAS.414.1793A}).  Assuming typical {\em un-extincted} (and age dependent) optical-to-$K$-band colour indices of 2 to 3\,mag, one would expect a corresponding bend to appear at $M_K \sim -12$ to $-13$\,mag.  \citet{2005A&A...443...41M} find a bend at a significantly brighter level $M_K \sim -15.5$\,mag in the Antennae (but see \citealp{2009ApJ...704..453F}) not unlike our result for IRAS\,18293-3413, suggesting significant extinction effects.
The bend-point of a double power-law at $M_K \sim -16.5$ mag, or the Schechter characteristic magnitude at $M_K^{\star} \sim -17.4$ mag, would correspond to a {\em minimum} characteristic mass of $M^{\star} \approx 1-2 \times 10^6 \, M_{\odot}$, or perhaps double this mass range if we adopt an average mass-to-light ratio over the first 30 Myr of age (see next Section).  These masses are higher by factors of 5 to 10 than the characteristic mass found for SSCs in normal spirals \citep{2006A&A...450..129G,2009A&A...494..539L} while they are very similar to those suggested for galaxies with much higher SFRs, such as LIRGs  \citep[see e.g.][]{2008MNRAS.390..759B}.
 However until there is clearer evidence for real bends in SSC LFs from a larger dataset it is premature to read too much into these masses.

\begin{figure}
\centering
\resizebox{1.1\hsize}{!}{\rotatebox{0}{\includegraphics{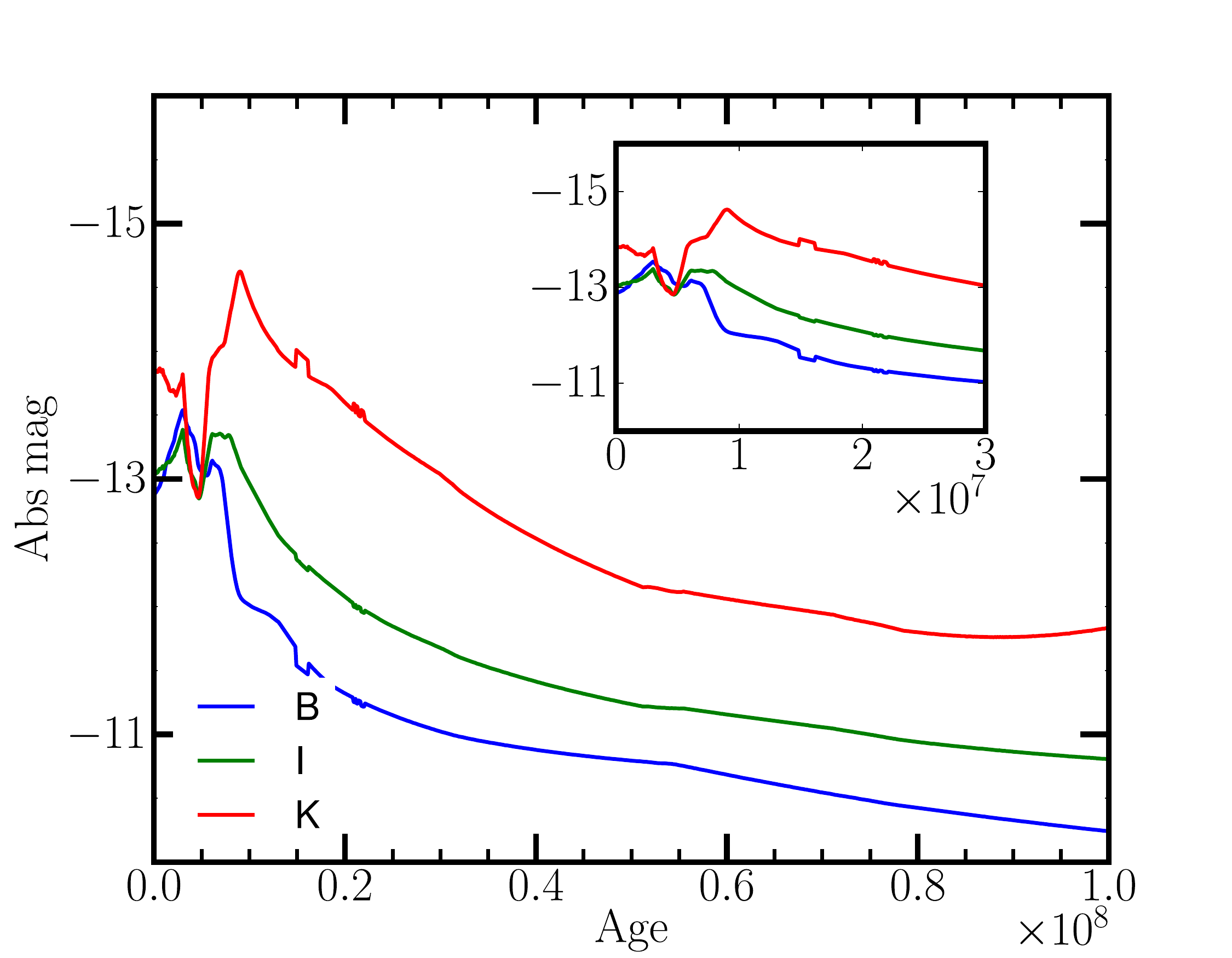}}}
\caption{\small {\tt Starburst99} model of 2\,$\times$\,$10^5\,M_\odot$\,SSP cluster with a Kroupa IMF. The three curves show the absolute magnitude of the star cluster in $B$-(blue), $I$-(green) and $K$-(red) bands. The inset highlights that the $K$-band luminosity of star clusters is likely to peak close to age 10\,Myr while they peak somewhat earlier in the optical.} 
\label{SB99}
\end{figure}

\subsection{Mass range approximations of the $K$-band SSCs}

An interesting characteristic of SSCs observed at NIR wavelenghts is a relatively narrow time-frame at ages  of $\sim$\,10\,Myr when the high mass stars in the SSCs enter the RSG phase making them very NIR-luminous and \textgravedbl suddenly\textacutedbl\,red in the optical-to-NIR colours compared to earlier blue stages.  This is seen in  Fig.\,\ref{SB99} which plots as an example the evolution of cluster brightness with time in the $BIK$ filters, derived from a {\tt Starburst99} \citep{1999ApJS..123....3L} model assuming an instantaneous SF with a  fixed mass of 2\,$\times$\,$10^5\,M_\odot$, and a  Kroupa IMF.  Hence, a {\em lower limit on SSC masses} detected in this work can be obtained by assuming the mass-to-light ratio at that $\sim 10$\,Myr age.   In many of our galaxies the most massive detected SSC has $M_K \sim -18$\,mag corresponding to a mass of $4 \times 10^6\,M_\odot$ at 10 Myr of age or more than  $2 \times 10^7\,M_\odot$ at 30 Myr; clearly we are sampling very massive clusters in LIRGs.  

SSCs with masses in excess of $10^7 \,M_\odot$ have been found before (e.g. \citealp{2006A&A...448..881B, 2010ARA&A..48..431P}). The faintest-detected SSCs in our sample have $M_K \sim -13$ corresponding to masses of $\approx 2 \times 10^4\,M_\odot$, while the photometric 80\,\% completeness limits in different targets correspond to lower limits of $1-2 \times 10^5\,M_\odot$.   We note that with such massive clusters, the luminosities of individual bright RSG stars are negligible compared to the total integrated flux of the star cluster. Therefore, stochastic effects should not introduce significant scatter in the mass-to-light ratios and inferred ages of our clusters \citep{ 2012ApJ...750...60F}.

\section{Summary}\label{summary}

In this work we studied the characteristics of massive star clusters in the extreme environments of local interacting LIRGs. Observations were performed in the $K$-band filter using two different NIR AO instruments with pixel scales of $\sim$\,0.02\textacutedbl${\rm pix}^{-1}$ and a FWHM\,$\sim$\,0.1\textacutedbl~for point sources. The galaxy sample consists of LIRGs in the redshift range 0.01\,$<$\,{\it z}\,$<$\,0.05.  This study is opening new territory in SSC studies since both the distance range and host galaxy SFR range are higher compared to most previous SSC studies.  Specifically, we derived the $K$-band luminosity functions of SSC candidates in our targets, and because of the distances involved,  we also carefully evaluated the effect of blending on the power-law index $\alpha$ of the LFs.  Extensive photometric completeness simulations were done, as well as checking the effects of sample binning and foreground contamination,  which turned out to not be significant.  The main results from this work can be summarised as:

\begin{enumerate}
\item{The SSC luminosity function is probed at high completion down to $M_K \sim -14$ or $-15$ mag in our sample.  In this range all the LFs are reasonably well fitted by a single power-law, though in the case of IRAS\,18293-3413 a double power-law  or a Schechter function is a better approximation.   The values of the best-fit slopes vary with a wide range from  $\alpha = 1.5$ to $2.4$ with the average value at $\alpha \approx 1.9$,  and the combined SSC LF, excluding the most distant target, at $\alpha \approx 1.8$.   The slopes appear slightly  flatter than those in normal spirals which typically have $\alpha \approx 2.2$.} The sample is still too small to find definitive trends of the slope with SFR or interaction stage.  One or more of age, extinction, and mass-dependent cluster disruption effects can all lead to small $\alpha$ values at the faint end, but cannot be unambiguously separated from the present dataset alone.

\item{We carefully examined the possibility of blending of SSCs in our target LIRGs.  Though blending does happen with the typical resolutions of 30 to 40\,pc physical sizes of our sample, we showed that it is not enough to change the LF slopes by more than $\Delta \alpha \approx 0.05-0.1$, nor change drastically the measured luminosities of SSCs.   Hence we conclude that out to $\sim 100$\,Mpc it is quite possible to accurately measure SSC properties with 8-m class telescope adaptive optics and $HST$ imaging.  In addition to deriving some general blending/confusion properties, we found that the photometric apertures used do not affect the LF slope, but small apertures are necessary to recover the luminosities as correctly as possible.  Worsening spatial resolution tends to flatten the measured LF slopes through increasing blending.  However, the effect becomes pronounced only when close to the confusion limit.  In the case of SSCs distributed as in the Antennae, we determined that LF slopes at $M_H < -12$ are not reliable if the spatial resolution corresponds to a physical size larger than $\sim 50$\,pc.}
\end{enumerate}

All our findings are based on observations with a single filter and thus the estimation of ages of the star cluster candidates was beyond the scope of this work. However, we are in the process of estimating the physical characteristics of the selected star cluster candidates with the help of archival $HST$ data and by using $J$ and $H$ imaging from Gemini  to be 
able to constrain the SSC evolution models more robustly. 
In addition, with a larger sample of southern LIRGs and starbursts with the VLT/NACO currently available, 
and the next generation of multi-conjugate AO systems promising to deliver much more stable PSFs across a larger field of view we expect to grow both the number and quality of SSC LFs to probe correlations with LIRG host galaxy characteristics and environments.

\section*{Acknowledgments}

We thank the two referees for their thoughtful and very useful comments. ZR acknowledges financial support from the South African Square Kilometre Array and PV from the National Research Foundation. SM and EK acknowledge funding from the Academy of Finland (project: 8120503). Based on observations obtained at the Gemini Observatory, which is operated by the 
    Association of Universities for Research in Astronomy, Inc., under a cooperative agreement 
    with the NSF on behalf of the Gemini partnership: the National Science Foundation 
    (United States), the National Research Council (Canada), CONICYT (Chile), the Australian 
    Research Council (Australia), Minist\'{e}rio da Ci\^{e}ncia, Tecnologia e Inova\c{c}\~{a}o 
    (Brazil) and Ministerio de Ciencia, Tecnolog\'{i}a e Innovaci\'{o}n Productiva (Argentina). 
    Observations were taken as part of programs GN-2008A-Q-38, GN-2008B-Q-32, GN-2009A-Q-12, GN-2009B-Q-23, and GN-2010A-Q-40 (PI: S.\,Ryder). And based in part on observations made with the European Southern Observatory telescopes, Paranal, Chile, under programmes 072.D-0433 and 073.D-0406. Based in part on observations made with the NASA/ESA $HST$, obtained from the data archive at the Space Telescope Science Institute, which is operated by the Association of Universities for Research in Astronomy, Inc., under NASA contract NAS 5-26555. Based in part on observations made with the Nordic Optical Telescope, operated on the island of La Palma jointly by Denmark, Finland, Iceland, Norway, and Sweden, in the Spanish Observatorio del Roque de los Muchachos of the Instituto de Astrofisica de Canarias.

%\bibliographystyle{apj}
%\small
%\bibliography{Nambib}

\end{document}